\documentclass[12pt]{article}

\usepackage{url}
\usepackage{bbm}
\usepackage{mathtools}
\usepackage{amssymb}
\usepackage{amsthm}
\usepackage{empheq}
\usepackage{latexsym}
\usepackage{enumitem}
\usepackage{eurosym}
\usepackage{dsfont}
\usepackage{appendix}
\usepackage{color} 
\usepackage[unicode]{hyperref}
\usepackage{frcursive}
\usepackage[utf8]{inputenc}
\usepackage[T1]{fontenc}
\usepackage{geometry}
\usepackage{multirow}
\usepackage{todonotes}
\usepackage{lmodern}
\usepackage{anyfontsize}
\usepackage{pgfplots}
\usepackage{stmaryrd}
\usepackage{float}
\usepackage{bookmark}

\graphicspath{ {images/} }

\pgfplotsset{compat=newest}

\definecolor{red}{rgb}{0.7,0.15,0.15}
\definecolor{green}{rgb}{0,0.5,0}
\definecolor{blue}{rgb}{0,0,0.7}
\hypersetup{colorlinks, linkcolor={red},citecolor={green}, urlcolor={blue}}
            
\makeatletter \@addtoreset{equation}{section}

\newtheorem{theorem}{Theorem}
\newtheorem{theorem2}{Theorem}[section]
\newtheorem{assumption}[theorem2]{Assumption}

\newtheorem{lemma}[theorem2]{Lemma}

\newtheorem{remark}[theorem2]{Remark}

\DeclareUnicodeCharacter{014D}{\=o}
\def \P{\mathbb{P}}

\def \R{\mathbb{R}}

\def\Dc{{\cal D}}

\def\Fc{{\cal F}}

\title{Market making and incentives design in the presence of a dark pool: a deep reinforcement learning approach\footnote{This work benefits from the financial support of the Chaires Analytics and Models for Regulation, Financial Risk, and Finance and Sustainable Development. The authors would like to thank Charles-Albert Lehalle for fruitful discussions and remarks. Bastien Baldacci and Mathieu Rosenbaum gratefully acknowledge the financial support of the ERC Grant 679836 Staqamof. Thibaut Mastrolia gratefully acknowledges the support of the ANR project PACMAN ANR-16-CE05-0027.}}
\author{Bastien {\sc Baldacci}\footnote{\'Ecole Polytechnique, CMAP, 91128, Palaiseau, France,  bastien.baldacci@polytechnique.edu.} \and Iuliia {\sc Manziuk}\footnote{Université Paris 1 Panthéon-Sorbonne. Centre d'Economie de la Sorbonne. 106, boulevard de l'Hôpital, 75013 Paris, France, iuliiamanziuk@gmail.com} \and Thibaut {\sc Mastrolia}\footnote{\'Ecole Polytechnique, CMAP, 91128, Palaiseau, France, thibaut.mastrolia@polytechnique.edu}\and 
Mathieu {\sc Rosenbaum}\footnote{\'Ecole Polytechnique, CMAP, 91128, Palaiseau, France, mathieu.rosenbaum@polytechnique.edu}}

\begin{document}

\maketitle
\begin{abstract}
We consider the issue of a market maker acting at the same time in the lit and dark pools of an exchange. The exchange wishes to establish a suitable make-take fees policy to attract transactions on its venues. We first solve the stochastic control problem of the market maker without the intervention of the exchange. Then we derive the equations defining the optimal contract to be set between the market maker and the exchange. This contract depends on the trading flows generated by the market maker's activity on the two venues. In both cases, we show existence and uniqueness, in the viscosity sense, of the solutions of the Hamilton-Jacobi-Bellman equations associated to the market maker and exchange's problems. We finally design deep reinforcement learning algorithms enabling us to approximate efficiently the optimal controls of the market maker and the optimal incentives to be provided by the exchange.\\

\noindent{\bf Keywords:} Market making, dark pools, regulation, make-take fees, stochastic control, principal-agent problem, deep reinforcement learning, actor-critic method

\end{abstract}

\section{Introduction}\label{Section Introduction}
Since the seminal work \cite{avellaneda2008high}, a vast literature on optimal market making problems has emerged. A market maker is a liquidity provider whose role is to post orders on the bid and ask sides of the limit order book of an underlying asset. Various extensions of \cite{avellaneda2008high} have been considered, see for example \cite{cartea2014buy,gueant2013dealing} and the books \cite{cartea2015algorithmic,gueant2016financial} for further references. In most of these works, it is assumed that there is no make-take fees system on the market. The problem of relevant make-take fees is studied quantitatively in \cite{baldacci2019optimal,el2018optimal}. In these papers, the policies are designed in the context of traditional liquidity venues, or so-called ``lit pools''. On these venues, the order book is visible to market participants, and transactions are fully transparent. Market takers can in particular monitor the quotes offered by market makers.\\

However, recent regulatory changes have induced a rise of different types of alternative trading mechanisms, notably ``dark pools'', which have gained a significant market share. Nowadays, many major exchanges, such as Bats-ChiX and Turquoise, have their dark pools in addition to their major trading platforms. Furthermore, several traditional exchanges such as NYSE and Euronext offer trading platforms whose functioning is inspired mainly by dark pools. Trading rules for dark pools are very diversified, but they share at least two important properties. The first one is the absence of a visible order book for market participants, which implies that investors have no information on the amount of liquidity posted by market makers. Second, aiming at improving prices for clients compared to the lit venue, dark pools usually set prices that are different from those in the lit pool. For example, many dark pools take the mid-price of the lit pool as their transaction price. Because of these two effects, it is presumed that trades in dark pools have no or less price impact.\footnote{Note however, that transactions' reporting imposed by regulation in most markets may still induce some delayed price impact.} This feature enables market makers to mitigate their inventory risk. Finally, a remarkable phenomenon is that dark pools are prone to a latency effect: the price being monitored in the lit pool can change between the time of a request in the dark pool and that of the corresponding transaction. Such price discrepancy due to latency is particularly frequent in the presence of high imbalance because the price is likely to move when liquidity is scarce on one side of the book.  \\

As the market impact of trades on a dark pool is less important or delayed, market makers can also use it to liquidate large positions. Therefore there is a trade-off between transacting in the dark pool at a lousy price with low impact or in the lit pool at a better price with higher impact. Dark pools are also very attractive for market takers because of the reduced market impact and the possibility to be executed at a better price than in the lit pool. \\

To our knowledge, most of studies treat the issue of trading in dark pools mainly from the point of view of optimal liquidation: a trader wishing to buy or sell a large number of shares of one or several stocks and needing to find an optimal order placement strategy between the lit and dark pools, see for example \cite{laruelle2011optimal}. In this paper, we rather focus on the behavior of a market maker, acting on both lit and dark venues. In the lit market, we assume that there is an efficient price $S_t$ and that the market maker always posts volumes on the bid and ask sides at prices $S_t + \frac{\mathcal{T}}{2}$ and $S_t - \frac{\mathcal{T}}{2}$, where $\frac{\mathcal{T}}{2}$ represents the half-tick of the market.\footnote{We have in mind here a large tick asset for which the spread equals the tick size.} The market maker also provides liquidity in the dark pool where the transaction price is the efficient price $S_t$ (possibly with the latency effect). This can partially be seen as the dual problem of \cite{el2018optimal}, without dark pools, where the posted volume is fixed at one unit, and the market maker optimizes the quoted spread. In addition to market impact and latency phenomena, we also take into account transaction costs for market orders on both venues, which can be smaller in the dark pool. Thus, in our setting, a single market maker only needs to select the volumes to post on the bid and ask sides of both lit and dark pools. \\

An exchange managing the lit and dark venues wishes to attract transactions. Inspired by the work~\cite{el2018optimal}, we consider that the exchange offers a contract to the market maker whose remuneration at a terminal time is determined according to the executed transactions on both venues. This is a so-called principal-agent framework, first formalized in \cite{cvitanic2016moral,cvitanic2018dynamic,sannikov2008continuous}. Here, the wealth of the exchange (the principal) depends on the market order flows, which are a function of the volumes posted by the market maker (the agent). However, the exchange cannot control those volumes and may only provide incentives to influence the market maker's behavior. These incentives take the form of a contract between the market maker and the exchange, whose payoff depends on observed trading flows. \\

To find an optimal contract and optimal volumes for the market maker in response to this contract, we need to solve a nonlinear Hamilton-Jacobi-Bellman (HJB for short) equation. Dimensionality~(above four) and complexity of the resulting equations do not allow us to apply classical root-finding algorithms. Therefore we use a method based on neural networks to solve our HJB equations. Neural networks have been at the core of recent studies on high-dimensional PDE resolution. In \cite{han2018solving}, the authors introduce a deep learning-based methodology that can handle general high-dimensional parabolic PDEs. This approach relies on the reformulation of PDEs via Backward Stochastic Differential Equations, where neural networks approximate the gradients of the unknown solution. Since then, many extensions have been proposed, see for example \cite{bachouch2018deep,langrenealgorithmes}.\\

In our setting, the market maker has to fix volumes in response to the incentives of the exchange. These volumes are functions of the incentives (and of the market maker's inventory), which are the solution of a nonlinear equation. The resolution of our principal-agent problem consists of two stages. The first stage is to represent the volumes posted by the market maker by a neural network. Taking into account the optimal response of the market maker to given incentives, the exchange needs to choose the contract maximizing its utility. So the second stage is to solve a HJB equation to obtain the optimal contract. However, dimensionality and the high degree of nonlinearity of this equation make standard numerical methods hard to apply. We circumvent this difficulty by adopting a reinforcement learning method. More precisely, we use an actor-critic approach where not only the controls of the exchange, but also its value function are represented by neural networks. The essence of this method is the alternation of the learning phases of the controls and of the value function. \\

The paper is organized as follows. Market dynamics are introduced in Section \ref{Section The Model}. In Section \ref{section market making without exchange}, we first investigate the problem of a market maker acting on both lit and dark venues without any incentive policy from the exchange. His goal is to maximize his PnL process while managing his inventory risk. It is a stochastic control problem, where the corresponding HJB equation cannot be solved explicitly. We show existence and uniqueness of a viscosity solution for this equation.\\

In Section \ref{section market making with exchange}, we analyze the bi-level optimization problem associated with the issue of optimal contracting between the market maker and the exchange owning both lit and dark pools. Following recent works on make-take fees policies mentioned above, we first prove a representation theorem for the contract proposed to the market maker. We then establish existence and uniqueness of a viscosity solution for the HJB equation corresponding to the problem of the exchange. \\

A key difference with \cite{baldacci2019optimal, el2018optimal} is the absence of a closed-form solution for the best response of the market maker to a given contract. Therefore, the HJB equation of the exchange cannot be solved explicitly. In Section~\ref{Section neural network}, we introduce a deep reinforcement learning method as a computational tool enabling us to address both exchange and market maker's problems in practice. We conclude this section with numerical experiments, illustrating various behaviors of the market maker under different market scenarios.

\section{The market model}\label{Section The Model}

\subsection{Stochastic basis}

The framework considered throughout this paper is inspired by the article \cite{avellaneda2008high} in which the authors investigate the problem of optimal market making without intervention of an exchange. Let $T>0$ be a finite horizon time and $\mathcal{V}^l,\mathcal{V}^d \subset \mathbb{N}$ the sets of possible values for volumes in the lit and dark pools, of cardinality $\#\mathcal{V}^l,\#\mathcal{V}^d$. We define $\Omega:=\Omega_{c} \times \Omega_{d}^{2(\#\mathcal{V}^l + \#\mathcal{V}^d)}$ with $\Omega_{c}$ the set of continuous functions from $[0,T]$ into $\mathbb{R}$ and $\Omega_{d}$ the set of piecewise constant c\`adl\`ag functions from $[0,T]$ into $\mathbb{N}$. $\Omega$ is a subspace of the Skorokhod space $\Dc([0,T],\R^{2(\#\mathcal{V}^l + \#\mathcal{V}^d)+1})$ of c\`adl\`ag functions from $[0,T]$ into $\R^{2(\#\mathcal{V}^l + \#\mathcal{V}^d)+1}$ and write $\Fc$ for the trace Borel $\sigma$-algebra on $\Omega$, where the topology is the one associated with the usual Skorokhod distance on $\Dc([0,T],\R^{2(\#\mathcal{V}^l + \#\mathcal{V}^d)+1})$. \\

We define $(\mathcal{X}_{t})_{t\in [0,T]}:=(W_{t},N_{t}^{i,j,k})_{t\in [0,T], i\in\{a,b\},j\in\{l,d\},k\in\mathcal{V}^j}$ as the canonical process on $\Omega$, that is for any $\omega:=(w,n^{i,j,k})\in \Omega$
\[
W_{t}(\omega):=w(t), \; N_{t}^{i,j,k}(\omega)=n^{i,j,k}(t), i\in \{a,b\}, j\in \{l,d\} \text{ and } k\in \mathcal{V}^j . 
\]
For any $i\in \{a,b\}$, $j\in \{l,d\}$ and $k\in\mathcal{V}^j$, $N_t^{i,j,k}$ denotes the total number of trades of size $k$ made between time $0$ and time $t$, where $a$, $b$ stand for the ask and bid side respectively and $l$, $d$ for the lit and dark pools respectively. Finally the process $W$ represents the mid-price of the traded asset. \\

Then we define the probability $\mathbb{P}^{0}$ on $(\Omega,\Fc)$ under which $W_t$ and the  $N_t^{i,j,k}$ are independent, $W_t$ is a one-dimensional Brownian motion and the $N_t^{i,j,k}, i\in \{a,b\}, j\in \{l,d\}, k\in\mathcal{V}^j$ are Poisson processes with intensity $\epsilon>0$ small enough.\footnote{In other words, $\P^0$ is the product measure of the Wiener measure on $\Omega_c$ and the unique measure on $\Omega_d^{2(\#\mathcal{V}^l+\#\mathcal{V}^d)}$ so that the canonical process corresponds to a multidimensional homogeneous Poisson process with arbitrary small intensity, representing a situation where no liquidity is available.} Finally, we endow the space $(\Omega,\mathcal{F})$ with the ($\P^0-$completed) canonical filtration $\mathbb{F}:=(\mathcal{F}_{t})_{t\in[0,T]}$ generated by $(\mathcal{X}_{t})_{t\in[0,T]}$. 

\subsection{Traded volumes, market impact and latency}

In this section, we formalize the connection between volumes posted by the market maker and arrival intensity of market orders on the ask and bid sides of both venues. We also take into account market impact phenomenon and latency effect in the dark pool. 

\subsubsection{Admissible controls, inventory process and market takers' arrival flows}\label{subbsection admissible controls and intensity}

Let $2\overline{q}\in\mathbb{N}$ represent a risk limit for the market maker, which corresponds to the maximum number of cumulated bid and ask orders the market maker can handle. We define the volume process $(\mathcal{L}_t)_{t\in [0,T]}:= (\mathcal{L}_t^l, \mathcal{L}_t^d)_{t\in [0,T]} \in(\mathcal{V}^l)^2\times(\mathcal{V}^d)^2$, where $\mathcal{L}_t^l = (\ell_t^{a,l},\ell_t^{b,l})_{t\in [0,T]}$ and $\mathcal{L}_t^d = (\ell_t^{a,d},\ell_t^{b,d})_{t\in [0,T]}$ with $\ell_{t}^{i,j}$ corresponding to the volume posted by the market maker at time $t$ on side $i\in\{a,b\}$ of pool $j\in\{l,d\}$. The set $\mathcal{A}$ of admissible controls of the market maker is therefore defined as
\begin{align*}
\mathcal{A}:=\big\{(\mathcal{L}_{t})_{t\in [0,T]} \text{ predictable, s.t for } i\in \{a,b\}, \ell^{i,l}+\ell^{i,d} \in [0,2\overline{q}] \big\}.   
\end{align*}
The market maker manages his inventory $Q_t$, defined as the aggregated sum of the volumes filled on both sides of the lit and dark pools, namely
\begin{align*}
Q_{t}:=\sum_{j\in \{l,d\}}\sum_{(k^{a,j},k^{b,j})\in(\mathcal{V}^j)^2} k^{b,j}N_{t}^{b,j,k}-k^{a,j}N_{t}^{a,j,k}.
\end{align*}
\begin{remark}
Note that we assume that there is no partial execution in our model. Therefore market orders consume the whole volume posted by the market maker on the considered side and pool.
\end{remark}

We define the function
\begin{align*}
\psi^{i,j}(\mathcal{L}^l_t):=  \left\{
    \begin{array}{ll}
        I^a(\mathcal{L}^l_t) \text{ if } (i,j)\in \{(a,l),(b,d)\}  \\
        I^b(\mathcal{L}^l_t) \text{ if } (i,j)\in \{(b,l),(a,d)\},
    \end{array}
\right.
\end{align*}
where $I^{a}(\mathcal{L}^l_t):=\frac{\ell_{t}^{a,l}}{\ell_{t}^{a,l}+\ell_{t}^{b,l}},I^{b}(\mathcal{L}^l_t):=\frac{\ell_{t}^{b,l}}{\ell_{t}^{a,l}+\ell_{t}^{b,l}}$ represent the imbalances on the ask and bid sides of the lit pool respectively. To model the behavior of market takers, we define the intensities of the processes~$N^{i,j,k}$ as
\begin{align*}
\lambda^{\mathcal L,i,j,k}_{t} := \lambda^{i,j}(\mathcal{L}_t^l)\mathbf{1}_{\{\phi(i)Q_{t^{-}} > -\overline{q},\ell_t^{i,j}=k\}}, \quad \phi(i) :=  \left\{
    \begin{array}{ll}
        1 \text{ if } i = a  \\
        -1 \text{ if } i = b,
    \end{array}
\right.
\end{align*}
where 
\begin{align*}
    \lambda^{i,j}(\mathcal{L}_t^l) := A^j \exp \Big(-\frac{\theta^j}{\sigma} \psi^{i,j}(\mathcal{L}^l_t)\Big)\mathbf{1}_{\{\mathcal{L}_t^{l}\neq (0,0)\}}+\epsilon\mathbf{1}_{\{\mathcal{L}_t^{l}= (0,0)\}},
\end{align*}
where $\sigma>0$ is the volatility of the asset's mid-price. A high imbalance on the ask side decreases the probability that an ask limit order is filled in the lit pool and conversely for the bid side. Moreover, when the imbalance on the ask (resp. bid) side of the lit pool is high, if a market taker wants to buy, it is worth trying it in the dark pool, because the high imbalance indicates that the ask price in the lit may not be competitive. The coefficients $\theta^l, \theta^d > 0$ represent the influence of the imbalance on the intensity of orders' arrivals and $A^l, A^d > 0$ are average order flow intensity parameters. \\

For $\mathcal{L} \in \mathcal{A}$, we introduce a new probability measure $\mathbb{P}^{\mathcal{L}}$ under which $W$ remains a one-dimensional Brownian motion and for $i\in \{a,b\}, j\in \{l,d\}$, $k\in\mathcal{V}^j$ the
\begin{align*}
& N_{t}^{\mathcal{L}, i, j, k}:=N_{t}^{i,j,k} - \int_{0}^{t} \lambda^{\mathcal L,i,j,k}_{u}\mathrm{d}u
\end{align*}
are martingales. This probability measure is defined by the corresponding Doléans-Dade exponential 
\begin{align*}
&L_{t}^{\mathcal{L}}:=\text{exp}\Bigg(\sum_{\substack{i\in \{a,b\} \\ j\in\{l,d\}}}\sum_{k\in \mathcal{V}^j}\int_{0}^{t}\mathbf{1}_{\{\phi(i)Q_{u^{-}}>-\overline{q},\ell_t^{i,j}=k\}}\bigg(\text{log}\Big(\frac{\lambda^{i,j}(\mathcal{L}^l_u)}{\epsilon}\Big)\mathrm{d}N_{u}^{i,j,k}-\Big(\lambda^{i,j}(\mathcal{L}^l_u)-\epsilon\Big)\mathrm{d}u\bigg)\Bigg),
\end{align*}
which is a true martingale by the uniform boundedness of the $\ell^{i,j}$.\footnote{The associated Novikov criterion is given in \cite{sokol2013optimal}.} We can therefore set the Girsanov change of measure with $\frac{\mathrm{d} \mathbb{P}^{\mathcal{L}}}{\mathrm{d} \mathbb{P}^{0}}|_{\mathcal{F}_{t}} = L_{t}^{\mathcal{L}}$ for all $t\in [0,T]$. In particular, all the probability measures $\mathbb{P}^{\mathcal{L}}$ indexed by $\mathcal{L}$ are equivalent. We write $\mathbb{E}^{\mathcal{L}}_{t}$ for the conditional expectation with respect to $\mathcal{F}_{t}$ under the probability measure $\mathbb{P}^{\mathcal{L}}$. We also define for $i\in\{a,b\},j\in\{l,d\}$ the processes 
\begin{align*}
    N_t^{i,j}:=\sum_{k\in \mathcal{V}^j}N_t^{i,j,k}
\end{align*}
of intensities $\lambda^{i,j}(\mathcal{L}_t^l)\mathbf{1}_{\{\phi(i)Q_{t^-}>-\overline{q}\}}$. These processes correspond to the total number of transactions executed on the bid or ask side of the lit or dark pools.

\subsubsection{Efficient price and market impact}

We define the efficient price of the underlying asset, observable by all market participants (in the sense that they can infer it) as
\begin{align*}
\tilde{S}_{t}:=\tilde{S}_{0}+\sigma W_{t},    
\end{align*}
where $\tilde S_0>0$ is the initial price of the underlying asset and $\sigma>0$ its volatility. When a limit order on the bid side is filled, the price decreases on average and conversely for the ask side (this is the so-called market impact, see for example \cite{bouchaud2010price, toth2017short}). Thus, we define the mid-price of the asset at time $t\in[0,T]$ by
\begin{align}\label{Impacted efficient price}
S_{t}:=\tilde{S}_{t}+\sum_{j\in \{l,d\}}\int_{0}^{t}\Gamma^{j}\ell_{u}^{a,j}\mathrm{d}N_{u}^{a,j}-\Gamma^{j}\ell_{u}^{b,j}\mathrm{d}N_{u}^{b,j},
\end{align}
where $\Gamma^{l},\Gamma^{d}>0$ are fixed constants representing the magnitude of market impact in the lit and dark pools. 

\begin{remark}
The market impact parameters $\Gamma^l, \Gamma^d$ are taken small enough with respect to the tick size to discard obvious arbitrage opportunities. Moreover, as the market impact in the dark pool is usually smaller or delayed compared to the lit pool, we will take $\Gamma^l\geq \Gamma^d$.
\end{remark}

\subsubsection{Latency in the dark pool}
We assume that in the lit pool, the best bid and best ask prices $P^{b,l}$ and $P^{a,l}$ satisfy
\begin{align*}
P_{t}^{b,l}:=S_{t}-\frac{\mathcal{T}}{2}, \; P_{t}^{a,l}:=S_{t}+\frac{\mathcal{T}}{2}, \; t\in[0,T],
\end{align*}
where $\frac{\mathcal{T}}{2}>0$ is the half tick of the market. In this setting, in the lit pool, the market maker only needs to control the volumes he posts. \\

In the dark pool, orders may be executed at the mid-price, which is a priori beneficial for market takers. In practice, due to latency effect in the dark pool, the mid-price can change by one half tick (or more) before the transaction is made. Therefore the order may be executed at a less advantageous price for the market taker (and sometimes at an even more advantageous one but we neglect this case for the sake of simplicity). Let us introduce the corresponding prices with and without latency: 
\begin{align*}
\begin{cases}
P_{t}^{b,d,\text{lat}} := S_{t}-\frac{\mathcal{T}}{2}, & P_{t}^{a,d,\text{lat}} := S_{t}+\frac{\mathcal{T}}{2}, \\
P_{t}^{b,d,\text{non-lat}}  :=S_{t}, & P_{t}^{a,d,\text{non-lat}}  :=S_{t}.
\end{cases}
\end{align*}
Recall that in most dark pools, market takers are supposed to be executed at the mid-price of the lit pool. However, the higher the imbalance on the ask (resp. bid) side of the lit pool, the higher the probability that the mid-price will move down (resp. up) quickly. To model the latency effect, we introduce Bernoulli random variables $\nu_{t}^{a}\sim \text{Ber}\big(I^{a}(\mathcal{L}^l_t)\big)$, $\nu_{t}^{b}\sim \text{Ber}\big(I^{b}(\mathcal{L}^l_t)\big)$ which are associated to each incoming market order in the dark pool.\footnote{We take the convention $I^{a}(0,0)=I^{b}(0,0)=0$.} If $\nu_t=1$, there is no latency, and conversely for $\nu_t = 0$. So we define
\begin{align*}
 N_{t}^{a,d,\text{lat}}:=\int_{0}^{t}(1-\nu_{u}^{a})dN^{a,d}_{u}, \quad N_{t}^{a,d,\text{non-lat}}:=\int_{0}^{t}\nu_{u}^{a}dN^{a,d}_{u},
\end{align*}
and 
\begin{align*}
 N_{t}^{b,d,\text{lat}}:=\int_{0}^{t}(1-\nu_{u}^{b})dN^{b,d}_{u}, \quad N_{t}^{b,d,\text{non-lat}}:=\int_{0}^{t}\nu_{u}^{b}dN^{b,d}_{u}.
\end{align*}
Note that for any $t\in [0,T]$, $N_{t}^{i,d,\text{lat}}+N_{t}^{i,d,\text{non-lat}}=N_{t}^{i,d}$ for $i\in \{a,b\}$. To our knowledge, our approach is the first one considering market making in the dark pool taking into account latency effect. 

\section{Market making without the intervention of the exchange}\label{section market making without exchange}

We address the problem of a market maker acting in the lit and dark pools, without intervention of the exchange. The profit and loss (PnL for short) of the market maker is defined as the sum of the cash earned from his executed orders and the value of his inventory. Thus it is expressed as
\begin{align*}
PL_{t}^{\mathcal{L}}:=\mathcal{W}_{t}^{\mathcal{L}}+Q_{t}S_{t},
\end{align*}
where, at time $t\in[0,T]$,
\begin{align*}
& \mathcal{W}_{t}^{\mathcal{L}}:=\int_{0}^{T}\Big(S_{t}+\frac{\mathcal{T}}{2}\Big)\ell_{t}^{a,l}\mathrm{d}N_{t}^{a,l}-\int_{0}^{T}\Big(S_{t}-\frac{\mathcal{T}}{2}\Big)\ell_{t}^{b,l}\mathrm{d}N_{t}^{b,l} +\int_{0}^{T}\Big(S_{t}+\frac{\mathcal{T}}{2}\Big)\ell_{t}^{a,d}\mathrm{d}N_{t}^{a,d,\text{lat}}\\
&\hspace{2em}+\int_{0}^{T}S_{t}\ell_{t}^{a,d}\mathrm{d}N_{t}^{a,d,\text{non-lat}}-\int_{0}^{T}\Big(S_{t}-\frac{\mathcal{T}}{2}\Big)\ell_{t}^{b,d}\mathrm{d}N_{t}^{b,d,\text{lat}}-\int_{0}^{T}S_{t}\ell_{t}^{b,d}\mathrm{d}N_{t}^{b,d,\text{non-lat}}
\end{align*}
represents his cash process and $Q_t S_t$ is the mark-to-market value of his inventory.\footnote{Note that for all $t\in[0,T]$, $\int_0^t\ell_u^{i,j}\mathrm{d}N_u^{i,j}=\sum_{k\in\mathcal{V}^j}k N_t^{i,j,k}$.} Note that market making activity in the dark pool without latency does not generate PnL through spread collection. We consider a risk averse market maker with exponential utility function and risk aversion parameter~$\gamma>~0$. We define his optimization problem as
\begin{align}\label{Market Maker's Problem}
V^{\text{MM}}_0=\sup_{\mathcal{L}\in \mathcal{A}}J_0^{\textup{MM}}(\mathcal{L}),
\end{align}
with for all $t\in [0,T]$,
\begin{align*} 
J_t^{\textup{MM}}(\mathcal{L})&= \mathbb{E}^{\mathcal{L}}\bigg[-\text{exp}\Big(-\gamma (PL_T^\mathcal{L}-PL_t^\mathcal{L})\Big)\bigg].
\end{align*}
Inspired by \cite{el2018optimal}, we prove a dynamic programming principle for the control problem \eqref{Market Maker's Problem}, see Section~\ref{Contract representation}, from which we derive the corresponding HJB equation. We define $\mathcal{O}=[0,2\overline{q}]^4$. Similarly to~\cite{gueant2013dealing}, we use a change of variable (see Equation \eqref{Change of variable Principal HJB} for the form of the ansatz) to reduce the initial problem to the following HJB equation:
\begin{align}\label{HJB equation market maker without exchange}
\begin{split}
&0=\partial_{t}v(t,q)+v(t,q)\frac{1}{2}\sigma^{2}\gamma^{2}q^{2}\\
& +\sup_{\mathcal{L}\in \mathcal{O}}\Bigg\{\!\!\sum_{(k^l,k^d)\in\mathcal{V}^l\times \mathcal{V}^d}\!\!\Bigg(\!\sum_{i\in \{a,b\}}  \lambda_t^{\mathcal{L},i,l,k^l}\!\bigg(\!\exp\!\Big(-\gamma \ell^{i,l}\big(\frac{\mathcal{T }}{2}+\Gamma^{l}(\phi(i)q-\ell^{i,l}) \big)\Big)v(t,q-\phi(i)k^{l})-v(t,q)\!\bigg) \\
&+ \!\!\!\!\sum_{i\in \{a,b\}}\!\sum_{\kappa\in K}\!\lambda_t^{\mathcal{L},i,d,k^d}\!\phi^{d}(i,\kappa)\bigg(\!\exp\!\Big(\!-\!\gamma \ell^{i,d}\big(\frac{\mathcal{T}}{2}\phi^{lat}\!(\kappa)\!+\!\Gamma^{d}(\phi(i)q\!-\!\ell^{i,d})\big)\Big)v(t,\!q\!-\phi(i)k^{d})\!-\! v(t,q)\bigg)\Bigg)\Bigg\},
\end{split}
\end{align}
with $K:=\{\text{lat}, \text{non-lat} \}$,
\begin{align*}
\phi^{\text{lat}}(\kappa) := \left\{
    \begin{array}{ll}
        1 \text{ if } \kappa= \text{ lat } \\
        0 \text{ if } \kappa= \text{ non-lat}
    \end{array}
\right. 
, \quad 
\phi^{d}(i,\kappa) := \left\{
    \begin{array}{ll}
         I^{b}(\mathcal{L}^l) \text{ if } (i,\kappa)\in \{(a,\text{lat}),(b,\text{non-lat})\} \\
         I^{a}(\mathcal{L}^l) \text{ if } (i,\kappa)\in \{(a,\text{non-lat}),(b,\text{lat}) \}
    \end{array}
\right.
\end{align*}
and terminal condition $v(T,\cdot)=-1$. We have the following theorem.

\begin{theorem}
There exists a unique viscosity solution to the HJB equation \eqref{HJB equation market maker without exchange}. It satisfies
\begin{align*}
    V_0^{\textup{MM}}=v(0,Q_0).
\end{align*}
The supremum in \eqref{HJB equation market maker without exchange} characterizes the optimal controls $\mathcal{L}^\star\in \mathcal{A}$.
\end{theorem}
The proof follows the same arguments as Theorem \ref{Main Theorem 2} in Section \ref{proof:visco}. \\  

We see that the supremum over $\mathcal{L}$ is not separable with respect to each control process as in \cite{el2018optimal, gueant2013dealing}. To our best knowledge there is no explicit expression for the optimal controls of the market maker. Nevertheless, as shown in Section \ref{section numerical results}, we can solve PDE \eqref{HJB equation market maker without exchange} numerically. More precisely, we make use of deep reinforcement learning techniques to approximate the optimal volumes posted of the market maker.

\section{Market making with the intervention of the exchange}\label{section market making with exchange}

Let us now consider the case where a make-take fees system is in place and influences the amount of liquidity provided by the market maker on both lit and dark venues.

\subsection{Modified PnL of the market maker}

Following the principal-agent approach of \cite{el2018optimal}, we now assume that the exchange gives to the market maker a compensation~$\xi$ defined as an $\mathcal{F}_{T}-$measurable random variable, which is added to his PnL process at terminal time $T$. This contract, designed by the exchange, aims at creating incentives so that the market maker attracts more transactions. \\

Therefore, the total payoff of the market maker at time $T$ is now given by $\mathcal W_T^{\mathcal{L}}+Q_TS_T+\xi$. The problem of the market maker then becomes 
\begin{align}\label{Market Maker's Problem with exchange}
V_0^{\textup{MM}}(\xi):=\sup_{\mathcal{L}\in \mathcal{A}}J_0^{\textup{MM}}(\mathcal{L},\xi),
\end{align}
with
\begin{align*} 
J_t^{\textup{MM}}\big(\mathcal{L},\xi\big)&:= \mathbb{E}_t^{\mathcal{L}}\bigg[-\text{exp}\Big(-\gamma (PL_T^\mathcal{L}-PL_t^\mathcal{L}+\xi)\Big)\bigg].
\end{align*}
To ensure that this functional is non-degenerate, we impose the following technical condition on $\xi$ (see the next section for the definition of an admissible contract):
\begin{align}
\sup_{\mathcal{L}\in \mathcal{A}}\mathbb{E}^{\mathcal{L}}\bigg[\text{exp}\Big(-\gamma'\xi\Big)\bigg]<+\infty, \text{ for some } \gamma'>\gamma,
\label{Integrability market maker}
\end{align}
so that the optimization problem of the market maker is well-posed. \\

For a fixed compensation $\xi$, the optimal response $\mathcal{L}^\star$ associated with the market maker's problem \eqref{Market Maker's Problem with exchange} is defined as
\begin{align*}
    \tag{OC}
    \label{OC}
    J_0^{\textup{MM}}(\mathcal{L}^\star,\xi)=V^{\text{MM}}_0(\xi) \quad \text{ for } \mathcal{L}^\star \in \mathcal{A}. 
\end{align*}

We now consider the problem of the exchange wishing to attract liquidity on its platforms. 

\subsection{Objective function of the exchange}

We assume that the exchange receives fixed fees $c^{l}, c^{d}>0$ for each market order occurring in the lit and dark pools respectively. As in \cite{el2018optimal}, since we are working on a short time interval, we take $c^{l},c^{d}$ independent of the price of the asset. \\

The goal of the exchange is essentially to maximize the total number of market orders sent during the period of interest. As the arrival intensities of market orders are controlled by the market maker through $\mathcal{L}$, the contract $\xi$ should aim at increasing these intensities. Thus, the exchange subsidizes the agent at time $T$ with the compensation $\xi$ so that its PnL is given by
\begin{align*}
\sum_{\substack{i\in\{a,b\} \\ j\in\{l,d\}}}c^{j}\int_{0}^{T}\ell^{i, j}_{t}\mathrm{d}N_{t}^{i,j}-\xi.
\end{align*}
We now need to specify the set of admissible contracts potentially offered by the exchange. We assume that the exchange has exponential utility function with risk aversion parameter $\eta>0$. The natural well-posedness condition for the problem of the exchange is
\begin{equation}\label{Integrability Principal}
\mathbb{E}^{\mathcal L^*}\bigg[\text{exp}\Big(\eta'\xi\Big)\bigg]<+\infty, \text{ for some } \eta'>\eta, 
\end{equation}
for any $\mathcal{L}^\star$ satisfying condition \eqref{OC}. \\

Since the $N^{i,j}$ are point processes with bounded intensities,  this condition, together with Hölder inequality, ensure that the problem of the exchange is well-defined. We also assume that the market maker only accepts contracts $\xi$ such that $V^{\text{MM}}_0(\xi)$ is above some threshold value $R<0$, that is $\xi$ must satisfy
\begin{align*}
    \tag{R}
    \label{(R)}
    \quad V^{\text{MM}}_0(\xi)\geq R. 
\end{align*} 
This threshold, called reservation utility of the agent, is the critical utility value under which the market maker has no interest in the contract. This quantity has to be taken into account carefully by the exchange when proposing a contract to the market maker. We can therefore define the space of admissible contracts $\mathcal{C}$ by
\begin{align*}
\mathcal{C}:=\Big\{ \xi \in \mathcal F_T,\text{ s.t }\text{\eqref{Integrability market maker}, \eqref{Integrability Principal} and \eqref{(R)} are satisfied} \Big\}.
\end{align*}
Thus the contracting problem the exchange has to solve is
\begin{align}
V_{0}^{E}:=\sup_{\xi \in \mathcal{C}}\mathbb{E}^{\mathcal{L}^\star}\Bigg[-\text{exp}\bigg(-\eta\Big(\sum_{\substack{i\in \{a,b\}\\ j\in\{l,d\}}}c^{j}\int_{0}^{T}\ell_{t}^{i,j}\mathrm{d}N_{t}^{i,j}-\xi\Big)\bigg)\Bigg].
\label{PnL Principal}
\end{align} 

In the next section, we characterize the form of an admissible contract $\xi\in\mathcal{C}$.\footnote{Note that for fixed $\xi$, the control $\mathcal{L}^\star$ is not necessarily unique. However, numerical results seem to indicate its uniqueness. Otherwise we could also consider a supremum over all $\mathcal{L}^\star$ satisfying \eqref{OC}, as it is usually done in principal-agent theory (see for instance \cite[Section 2.4]{cvitanic2018dynamic}).} 

\subsection{Design of an optimal make-take fees policy }\label{subsection design optimal contract}
\subsubsection{A class of contracts built on transactions}
Inspired by \cite{el2018optimal}, we prove in this section that without loss of generality, we can consider a specific form of contracts, defined by some $Y_{0}\in \mathbb{R}$ and a predictable process $Z=(Z^{\tilde{S}},Z^{i,j,k})_{i\in\{a,b\},j\in\{l,d\},k\in\mathcal{V}^j}$ chosen by the principal. A contract $\xi$ of this form can be written as
\begin{align}\label{Process Yt}
& \xi \!=\! Y_{T}^{Y_{0},Z}\!\!:=Y_{0}+\!\!\int_{0}^{T}\Big(\sum_{\substack{i\in \{a,b\} \\ j\in\{l,d\}}}\sum_{k\in \mathcal{V}^j}Z_{u}^{i,j,k}\mathrm{d}N_{u}^{i,j,k}\Big)\!+\! Z_{u}^{\tilde{S}}\mathrm{d}\tilde{S}_{u}+\Big(\frac{1}{2}\gamma\sigma^{2}(Z_{u}^{\tilde{S}}+Q_{u})^{2}\!-\! H(Z_{u},Q_{u})\Big)\mathrm{d}u,
\end{align}
where
\begin{align}\label{Definition H}
H(z,q):=\sup_{\mathcal{L}\in \mathcal{A}}h(\mathcal{L},z,q),
\end{align}
and $h:\mathbb{R}_+^{4}\times\mathbb{R}^{2(\#\mathcal{V}^l+\#\mathcal{V}^d)}\times\mathbb{Z}\rightarrow \mathbb{R}$ is the Hamiltonian of the agent's problem.\footnote{Its form is defined in \eqref{Hamiltonian Agent}. This Hamiltonian term appears naturally when applying the dynamic programming principle for the market maker's problem.}
To ensure admissibility of the contract, the process $(Z_t)_{t\in[0,T]}$ has to satisfy the following technical conditions:
 \begin{align}
\sup_{\mathcal{L}\in \mathcal{A}}\sup_{t\in [0,T]}\mathbb{E}^{\mathcal{L}}\bigg[\text{exp}\Big(-\gamma' Y_{t}^{0,Z}\Big)\bigg]< +\infty, \text{ for some } \gamma'>\gamma,
\label{Strong integrability MM}
\end{align}
and  
\begin{align}\label{First condition admissibility contracts}
\int_{0}^{T}|Z_{t}^{\tilde{S}}|^{2}+|H(Z_{t},Q_{t})|\mathrm{d}t < +\infty.
\end{align}
Given this integrability condition, the process $(Y_{t}^{0,Z})_{t\in [0,T]}$ is well-defined. The contract consists of the following elements:
\begin{itemize}
 \item The constant $Y_0$ is calibrated by the exchange to ensure that the reservation utility constraint \eqref{(R)} of the market maker is satisfied.\footnote{From Theorem \ref{Theorem Market Maker}, $\hat{Y}_0=-\gamma^{-1}\log(-R)$ ensures that the reservation utility constraint of the market maker is satisfied. }
 \item The term $Z^{\tilde{S}}$ is the compensation given to the market maker with respect to the volatility risk induced by the efficient price $\tilde{S}$.
 \item Every time a trade of size $k$ occurs on the ask or bid side of the lit or dark pool, the market maker receives $Z^{i,j,k}$.
 \item The term $\frac{1}{2}\gamma\sigma^{2}(Z^{\tilde{S}}+Q)^{2}-H(Z,Q)$ is a continuous coupon given to the market maker. 
\end{itemize}
\begin{remark}
In our setting, the volumes of limit orders do not belong to the canonical process and so the principal does not contract on the volumes displayed by the market maker. It is very reasonable as in practice, a large part of volumes sent by market makers are not executed or rapidly canceled. Therefore it is clearly preferable to build contracts based on actual transactions. Moreover, note that $\tilde{S}$, and not the mid-price $S$, appears in the contract \eqref{Impacted efficient price}. This is not an issue since $S$ can be decomposed into elements of the canonical process.
\end{remark}

Formally stated, the definition of the space $\Xi$ of contracts of the form \eqref{Process Yt} is
\[ \Xi\!=\! \big\{ Y_T^{Y_0,Z}\!\in\! \mathcal F_T, Y_{0}\!\in\!\mathbb R, Z\in \mathcal{Z},\; \text{s.t } \text{\eqref{(R)} holds} \big\},\]
where $\mathcal Z$ denotes the set of processes defined by
\begin{align}\label{def:Zespace}
\mathcal{Z}:=\big\{(Z^{\tilde{S}},Z^{i,j,k}), i\in \{a,b\},j\in \{l,d\}, k\in\mathcal{V}^j \text{ s.t } \eqref{Integrability Principal}, \eqref{Strong integrability MM}, \eqref{First condition admissibility contracts} \text{ are satisfied} \big\}.
\end{align}

\subsubsection{Solving the market maker's problem}\label{subsection Main theorem 1}

For $(\mathcal{L},z,q)\in (\mathcal{V}^l)^2\times(\mathcal{V}^d)^2 \times\mathbb{R}^{2(\#\mathcal{V}^l+\#\mathcal{V}^d)}\times\mathbb{Z}$ we define the Hamiltonian of the market maker, which appears in the contract $Y_T^{Y_0,Z}$ via the continuous coupon $H(Z,Q)$, by
\begin{align}
\begin{split}
h(\mathcal{L}&,z,q)\!:=\!\!\!\sum_{(k^l,k^d)\in\mathcal{V}^l\times \mathcal{V}^d} \Bigg(\sum_{i\in\{a,b\}}\; \!\!\gamma^{-1}\Bigg(\!\bigg(1\!-\!\text{exp}\Big(\!-\!\gamma\big(z^{i,l,k^l}+\ell^{i,l}(\frac{\mathcal{T }}{2}\!+\!\phi(i) \Gamma^{l}q)\!-\!\Gamma^{l}(\ell^{i,l})^{2} \big)\!\Big)\!\bigg)\lambda^{\mathcal L, i,l,k^l}_t\\
& \!\!\!+\!\!\sum_{\kappa\in K}\!\bigg(\!1\!-\!\text{exp}\Big(\!\!-\!\gamma\big(z^{i,d,k^d}\!+\!\ell^{i,d}(\frac{\mathcal{T}}{2}\!\phi^{\text{lat}}(\kappa)\!+\!\phi(i)\Gamma^{d}q)\!-\!\Gamma^{d}(\ell^{i,d})^{2} \big)\!\Big)\!\bigg)\lambda^{\mathcal L, i,d,k^d}_t\phi^{d}(i,\kappa)\Bigg)\Bigg).
\end{split}
\label{Hamiltonian Agent}
\end{align}

The next theorem states that the two sets $\mathcal{C}$ and $\Xi$ are, in fact, equal. Moreover, the contract representation \eqref{Process Yt} enables us to provide a solution to the market maker's problem \eqref{Market Maker's Problem with exchange}. The proof is given in Section \ref{section:preuveagent}.

\begin{theorem}\label{Theorem Market Maker}
Any admissible contract can be written under the form \eqref{Process Yt}, that is $\mathcal{C}=\Xi$. Moreover, for any $Y_{T}^{Y_{0},Z} \in \Xi$ we have \begin{align*}
V_0^{\textup{MM}}(Y_{T}^{Y_{0},Z})=-\exp(-\gamma Y_{0}),
\end{align*}
and Condition \eqref{OC} with $\xi = Y_{T}^{Y_{0},Z}$ is equivalent to the fact that $\mathcal{L}$ satisfies $h(\mathcal{L}_t,Z_t,Q_t)=H(Z_{t},Q_{t})$ for any $t\in [0,T]$. 
\end{theorem}

This theorem provides a tractable form of contracts for the design of a suitable make-take fees policy. \\

Given the knowledge of the market maker's response to a given contract, we reformulate the problem of the exchange and prove the existence and uniqueness of the associated value function. 

\subsection{Problem of the exchange}\label{subsection solving exchange problem}
\subsubsection{Reformulation of the problem}
Following Theorem \ref{Theorem Market Maker}, the contracting problem \eqref{PnL Principal} is reduced to
\begin{align}
V_{0}^{\text{E}}:= \underset{h(\mathcal{L}_t,Z_t,Q_t) = H(Z_{t},Q_{t}), \forall t\in [0,T]}{\underset{(Y_{0},Z,\mathcal L) \in \mathbb{R}\times \mathcal{Z}\times \mathcal A, }{\sup}}  \mathbb{E}^{\mathcal{L}}\Bigg[-\text{exp}\bigg(-\eta\Big(\sum_{\substack{i=\{a,b\} \\ j=\{l,d\}}}c^{j}\int_{0}^{T}\ell_{t}^{i,j}\mathrm{d}N_{t}^{i,j}-Y_{T}^{Y_{0},Z}\Big)\bigg)\Bigg].
\label{Principal Problem moral hazard}
\end{align} 

For a given contract $Y^{Y_0, Z}$, due to the form of \eqref{Hamiltonian Agent}, the market maker's optimal response does not depend on $Y_{0}$. With the exchange's objective function being decreasing in $Y_{0}$, the maximization with respect to $Y_0$ is achieved at the level $\hat{Y}_{0} = -\gamma^{-1}\log(-R)$. Therefore Problem \eqref{Principal Problem moral hazard} can be reduced to
\begin{align}\label{Control problem Principal for HJB}
v_{0}^{E}:=\underset{  h(\mathcal{L}_t,Z_t,Q_t)=H(Z_{t},Q_{t}), \forall t\in [0,T] }{\underset{(Z,\mathcal L) \in  \mathcal{Z}\times \mathcal{A} }{\text{sup }}} \mathcal J(Z,\mathcal L),
\end{align}
where 
\[\mathcal J(Z,\mathcal L)= \mathbb{E}^{\mathcal{L}}\Bigg[-\text{exp}\bigg(-\eta\Big(\sum_{\substack{i=\{a,b\} \\ j=\{l,d\}}}c^{j}\int_{0}^{T}\ell_{t}^{i,j}\mathrm{d}N_{t}^{i,j}-Y_{T}^{0,Z}\Big)\bigg)\Bigg].\]

\subsubsection{A bi-level optimization problem}
We define $\mathfrak D := [0,T]\times \mathbb R\times \mathbb N^{2(\#\mathcal{V}^l+\#\mathcal{V}^d)}\times \mathbb N^{2(\#\mathcal{V}^l+\#\mathcal{V}^d)} \times \mathbb R$ and for any vector $p$, $i\in \{1,\dots,\#p\}$, $p^{-i}=(p^1,\dots,p^{i-1},p^{i+1},\dots,p^{\#p})\in\mathbb{N}^{\#p-1}$. By Equation \eqref{PnL Principal} and the corresponding footnote, there might be more than one optimal response $\mathcal{L}^\star$ of the market maker. We show here how to solve the principal's problem for a specific optimal response $\mathcal{L}^\star$.\footnote{If there are several optimal responses $\mathcal{L}^\star$, the exchange should solve the HJB equation \eqref{HJB Principal before change variable} for every $\mathcal{L}^\star$ and, according to principal-agent theory, choose the optimal response that maximizes its own utility.} Using a dynamic programming principle similar to the one in Lemma \ref{Lemma 6.2}, we write the value function of the exchange's problem, $v^{E}: \mathfrak D \to \mathbb{R}$, as
\begin{align*}
v^{E}(t,\tilde{S}_t,\bar{N}_t,N_t,Y_t):=\underset{Z \in \mathcal{Z} }{\text{sup }} \mathbb{E}_t^{\mathcal{L}^\star}\Bigg[-\text{exp}\bigg(-\eta\Big(\sum_{\substack{i=\{a,b\} \\ j=\{l,d\}}}\sum_{k\in\mathcal{V}^j}c^{j}(\bar{N}_T^{i,j,k}-\bar{N}_t^{i,j,k})-Y_{T}^{0,Z}\Big)\bigg)\Bigg],
\end{align*} 
with 
\begin{align*}
(\bar{N}_t,N_t):=\big( k N_t^{i,j,k},N_t^{i,j,k}\big)_{i=\{a,b\}, j=\{l,d\}, k\in\mathcal{V}^j},
\end{align*}
and $\mathcal{L}^\star = (\ell^{\star b, l}(z, q), \ell^{\star a, l}(z, q), \ell^{\star b, d}(z, q), \ell^{\star a, d}(z, q))$ the optimal response of the market maker, in the sense of \eqref{Definition H}, displayed at time $t$ for a given inventory $q$ and given incentives $z$ of the exchange. \\

Recall that $Q_t= \sum_{j\in\{l,d\}}\sum_{k\in\mathcal{V}^j}(\bar N_t^{b,j,k} - \bar{N}_t^{a,j,k})$. Usual arguments enables us to show that $v^E$ is a viscosity solution of the HJB equation defined on $\mathfrak{D}$ by
\begin{align}\label{HJB Principal before change variable}
\begin{split}
    0 = &\partial_t v^E + \frac{1}{2}\sigma^2\partial_{\tilde{S}\tilde{S}}v^E +\sup_{z^{\tilde{S}}\in \mathbb{R}} \frac{\gamma\sigma^2}{2}(z^{\tilde{S}}+q)^2\partial_y v^E + \frac{\sigma^2}{2}(z^{\tilde{S}})^2\partial_{yy}v^E+\sigma^2 z^{\tilde{S}}\partial_{\tilde{S}y}v^E \\
    & +\sup_{z\in \mathbb{R}^{2(\#\mathcal{V}^l+\#\mathcal{V}^d)}} \sum_{\substack{i=\{a,b\} \\ j=\{l,d\}}}\sum_{k\in\mathcal{V}^j}\lambda^{\mathcal{L}^\star, i,j,k}_t\Big(\Delta_{i,j,k}(z)v^E-\partial_y v^E \mathcal{E}(z^{i,j,k},\ell^{\star i,j}(z,q))\Big),
\end{split}
\end{align}
where, for $i\in\{a,b\},j\in \{l,d\}$,
\begin{align*}
   & \Delta_{i,j,k}(z)v^E(t,\tilde{s},\bar{n},n,y):= v^E(t,\tilde{s},\bar{n}^{i,j,k}+k,\bar{n}^{-(i,j,k)},n^{i,j,k}+1,n^{-(i,j,k)},y+z^{i,j,k})-v^E(t,\tilde{s},\bar{n},n,y),\\
   &\mathcal{E}(z^{i,l,k},\ell^{\star i,l}(z,q)):=\frac{1}{\gamma}\Bigg(1-\textup{exp}\Big(-\gamma\big(z^{i,l,k}+\ell^{\star i,l}(z,q)(\frac{\mathcal{T}}{2}+\phi(i) \Gamma^{l}q)-\Gamma^{l}(\ell^{\star i,l}(z,q))^{2} \big)\Big)\Bigg), \\
   & \mathcal{E}(z^{i,d,k},\ell^{\star i,d}(z, q)):=\frac{1}{\gamma}\sum_{\kappa\in K}1\!-\!\textup{exp}\Big(\!-\!\gamma\big(z^{i,d,k}\!+\!\ell^{\star i,d}(z,q)(\frac{\mathcal{T}}{2} \phi^{\text{lat}}(\kappa)\!+\!\phi(i)\Gamma^{d}q)\!-\!\Gamma^{d}(\ell^{\star i,d}(z,q))^{2} \big)\Big)\phi^{d}(i,\kappa)\Bigg),
\end{align*}
and terminal condition 
\begin{align*}
    v^E(T,\tilde{s},\bar{n},n,y)=-\exp\big(-\eta(\sum_{\substack{i=\{a,b\} \\ j=\{l,d\}}}\sum_{k\in\mathcal{V}^j}c^j \bar{n}^{i,j,k}-y)\big).
\end{align*}
Remark that the best response of the market maker, for which we do not have explicit expression, appears in the value function of the exchange. Inspired by \cite{el2018optimal, gueant2013dealing}, we use the following ansatz for Equation \eqref{HJB Principal before change variable}:
\begin{align}\label{Change of variable Principal HJB}
    v^{E}(t,s,\bar{n},n,y)=v(t,q)\exp\big(-\eta(\sum_{\substack{i=\{a,b\} \\ j=\{l,d\}}}\sum_{k\in\mathcal{V}^j}c^j \bar{n}^{i,j,k}-y)\big),
\end{align}
where $v$ is a solution of the following HJB equation
\begin{align}\label{HJB Equation Principal First Form}
\left\{
    \begin{array}{ll}
        0=\partial_{t}v(t,q) + \mathcal{H}\big(q,\mathcal{L}^\star,v(t,\cdot)\big), \quad q\in\{-\overline{q},\dots,\overline{q}\},t\in[0,T),  \\
        v(T,q)=-1,
    \end{array}
\right.
\end{align}
with
\begin{align}
\label{hamiltonian}
\mathcal{H}\big(q,\mathcal{L}^\star_t,v(t,\cdot)\big)\!:=\! \sup_{z\in \mathbb{R}^{2(\#\mathcal{V}^l+\#\mathcal{V}^d)+1}}\mathcal{U}\big(z,q,\mathcal{L}^\star(z,q),v(t,\cdot)\big),
\end{align}
and
\begin{align*}
& \mathcal{U}\big(z,q,\mathcal{L}^\star(z,q),v(t,\cdot)\big)  := v(t,q)\Big(\frac{\eta}{2}\sigma^{2}\gamma\big(z^{\tilde{S}}+q\big)^{2}+\frac{\eta^{2}\sigma^{2}}{2}\big(z^{\tilde{S}}\big)^{2}\Big)  \\ 
& + \sum_{\substack{i=\{a,b\} \\ j=\{l,d\}}}\sum_{k\in\mathcal{V}^j} \lambda_t^{\mathcal{L}^\star,i,j,k} \bigg(\exp({\eta(z^{i,j,k}-k c^{j})})v\big(t,q-\phi(i) k\big)-v\big(t,q\big)\big(1+\eta\; \mathcal{E}(z^{i,j,k},\ell^{\star i,j}(z,q))\big)\bigg).
\end{align*}
This ansatz leads to dimensionality reduction from five to two parameters. Using \cite[Corollary 1.4.2]{bouchard2007introduction}, there exists a unique continuous viscosity solution associated to \eqref{HJB Equation Principal First Form}. 

\begin{remark}
Note that the supremum over $z^{\tilde{S}}$ is explicit and given by $z^{\tilde{S}}=-\frac{\gamma}{\gamma+\eta}q$ as in \cite{el2018optimal}. 
\end{remark}

\subsubsection{Solving the exchange's problem}

Making use of the ansatz $v$, the bi-level optimization problem \eqref{Control problem Principal for HJB} is reduced to solving the following system:
\begin{align}\label{Definition bilevel optimization problem}
\left\{
    \begin{array}{ll}
        0=\partial_{t}v(t,q) + \mathcal{H}\big(q,\mathcal{L}^\star_t,v(t,\cdot)\big), \quad \text{ with final condition } v(T,q)=-1,  \\
        h(\mathcal{L}^\star,z,q)=H(z,q), \quad q\in\{-\overline{q},\dots,\overline{q}\}.
    \end{array}
\right.  
\end{align}
We have the following theorem.
\begin{theorem}\label{Main Theorem 2}
There exists a unique continuous viscosity solution to HJB equation \eqref{HJB Equation Principal First Form}. It satisfies 
\begin{align*}
    v_0^E=v(0,Q_0)=v^E(0,\tilde{S}_0,\bar{N}_0,N_0,Y_0).
\end{align*}
Moreover, the optimal incentives of the principal $Z^\star$ are solutions of the supremum in \eqref{HJB Equation Principal First Form}.
\end{theorem}
The proof can be found in Section \ref{proof:visco}. \\

Theorem \ref{Main Theorem 2} allows us to use numerical methods to obtain the optimizers 
\begin{align}\label{Unique solution bilevel problem}
\Big(Z^\star(t,Q_{t^-}),\mathcal{L}^\star\big(Z^\star(t,Q_{t^-}),Q_{t^-}\big)\Big)_{t\in[0,T]}    
\end{align}
of the bi-level problem \eqref{Definition bilevel optimization problem}. Moreover, the optimal contract is given by
\begin{align*}
    \xi^\star\!=\hat{Y}_{0}+\!\!\int_{0}^{T}\Big(\sum_{\substack{i\in \{a,b\} \\ j\in\{l,d\}}}\sum_{k\in \mathcal{V}^j}Z_{u}^{\star i,j,k}\mathrm{d}N_{u}^{i,j,k}\Big)\!+\! Z_{u}^{\star \tilde{S}}\mathrm{d}\tilde{S}_{u}+\Big(\frac{1}{2}\gamma\sigma^{2}(Z_{u}^{\star \tilde{S}}+Q_{u})^{2}\!-\! H(Z^{\star}_{u},Q_{u})\Big)\mathrm{d}u.
\end{align*}

The second problem in \eqref{Definition bilevel optimization problem} is a classical optimization problem. Having found numerically $\mathcal{L}^\star(z,q)$, we solve the Hamilton-Jacobi-Bellman \eqref{HJB Equation Principal First Form} using neural networks. 

\begin{remark}
Theorem \ref{Main Theorem 2} characterizes only the value function of the exchange and not the optimal incentives defined in \eqref{Unique solution bilevel problem}, which are computed through deep reinforcement learning techniques. In particular, there is no guarantee of admissibility of the incentives $(Z^\star (t,Q_{t^-}))_{t\in[0,T]}$ solving~\eqref{hamiltonian}. However, we observe numerically (see Figure \ref{Optimal incentives bench}) that these incentive parameters are essentially linear (despite nonlinear nature of neural networks) in the inventory $Q$ at any fixed time $t$. This result is indeed quite usual in the optimal market making literature where asymptotic development of the function $v$ is used, so $v$ should be regular enough (see \cite[Section 4]{gueant2013dealing} or~\cite[Section~3.2]{avellaneda2008high}). The linearity of the incentives $Z^\star$ implies them to be in the set of admissible contracts $\mathcal Z$ defined by \eqref{def:Zespace}.
%
%
\end{remark}

\section{Numerical solution: a deep reinforcement learning approach}\label{Section neural network}

We now turn to the description of our numerical method to solve \eqref{Definition bilevel optimization problem}, the optimization procedure consists of two stages. At the first stage, we optimize the controls of the market maker for all possible values of the incentives given by the exchange. At the second stage, we use an actor-critic approach, to obtain both the optimal controls and the value function of the exchange. We conclude this section with numerical experiments showing the impact of incentives as well as that of market conditions on the volumes posted by the market maker on both lit and dark venues regulated by the exchange. Throughout these experiments, we assume the following:

\begin{assumption}
For all $i\in\{a,b\},j\in\{l,d\}$ and $k\in\mathcal{V}^j$, $Z^{i,j,k}=Z^{i,j}\in\mathbb{R}$.
\end{assumption}
This means that the principal provides incentives only with respect to the number of transactions on each side of each pool independently of the volumes. In that case, HJB equation \eqref{HJB Equation Principal First Form} remains valid. Recall that the optimal incentives depend on time and market maker's inventory, therefore, implicitly they depend on the transacted volume.\\

There is obvious bid-ask symmetry in our model with respect to the inventory of the market maker, as it can be seen in Hamiltonian \eqref{Hamiltonian Agent}. Thus for our numerical experiments we impose symmetry of the incentives with respect to $q$. As a consequence, we have symmetry of volumes posted by the market maker with respect to $q$, given incentives satisfying the bid-ask symmetry property. 

\subsection{Description}
\subsubsection{Market maker's problem}
The first step to tackle our principal-agent problem is to find optimal volumes $\mathcal{L}^\star = (\ell^{\star a, l}, \ell^{\star b, l}, \ell^{\star a, d}, \ell^{\star b, d})$, by solving for any couple $(z,q)$, the maximization problem of the market maker \eqref{Definition H}. To do so, we introduce a continuous version of the Hamiltonian \eqref{Hamiltonian Agent} with respect to $\mathcal L$, that is we maximize the following functional:
\begin{align}
\begin{split}
\mathcal L\longmapsto h^{c}(\mathcal{L}&,z,q)\!:=\!\sum_{i\in\{a,b\}}\; \!\!\gamma^{-1}\Bigg(\!\bigg(1\!-\!\text{exp}\Big(\!-\!\gamma\big(z^{i,l}+\ell^{i,l}(\frac{\mathcal{T }}{2}\!+\!\phi(i) \Gamma^{l}q)\!-\!\Gamma^{l}(\ell^{i,l})^{2} \big)\!\Big)\!\bigg)\lambda^{i,l}(\mathcal{L}_t^l)\\
& \!\!\!+\!\!\sum_{\kappa\in K}\!\bigg(\!1\!-\!\text{exp}\Big(\!\!-\!\gamma\big(z^{i,d}\!+\!\ell^{i,d}(\frac{\mathcal{T}}{2}\!\phi^{\text{lat}}(\kappa)\!+\!\phi(i)\Gamma^{d}q)\!-\!\Gamma^{d}(\ell^{i,d})^{2} \big)\!\Big)\!\bigg)\lambda^{i,d}(\mathcal{L}_t^l)\phi^{d}(i,\kappa)\Bigg).
\end{split}
\label{Hamiltonian Agent continuous version}
\end{align}
For fixed incentives, we have that $\mathcal L^{\star a,l}(q)=\mathcal L^{\star b,l}(-q)$. Because of the intricate form of the function $h^c$, we cannot have an explicit solution to the first order condition $\nabla_\mathcal{L}h^c = 0$, which is four-dimensional. Moreover, we do not have an \textit{a priori} knowledge on the functional form of optimizers $\mathcal{L}^\star:  \mathbb{R}^4\times [-\overline{q},\overline{q}] \to \mathbb{R}_+^4$, so we cannot apply canonical root-finding methods. Therefore to address this problem, we approximate the best response of the market maker by a neural network. \\

Although we do not use a purely grid-based method, we need to define a domain for arguments $q$ and $z$. In our model inventory $q$ of the market maker is bounded and evolves between risk limits $-\overline{q}$ and $\overline{q}$. We also define a bound $\overline{z}$ for the incentives $z \in \mathbb{R}^4$, so that $z \in [-\overline{z}, \overline{z}]^4$. This is in fact justified also by the paper \cite{el2018optimal} in which optimal incentives are proved to be bounded. \\ 

We approximate the best response function $\mathcal L^\star$ by a neural network $l[\omega^l]$, where $\omega^l$ are the weights of the neural network.\footnote{Here we slightly abuse notation denoting by $l[\omega^l]$ the response of the market maker obtained via neural network parametrized by weights $\omega^l$.} The neural network $l[\omega^l]$ takes as inputs principal's incentives and the market maker's current inventory $(z^{a,l}, z^{b,l}, z^{a,d}, z^{b,d}, q)$, which are normalized by $\overline{z}$ and $\overline{q}$ respectively. The network is composed of $2$ hidden layers with 10 nodes in each of them and with ELU activation functions. ELU activation function is of the form
\begin{align*}
    \textrm{ELU}(x) = \begin{cases}
    \alpha(e^x-1),\text{ for $x \le 0$}\\
    x, \text{ for $x > 0$},
    \end{cases}
\end{align*}
where $\alpha$ is a non-negative parameter, usually taken equal to $1$. \\

The final layer of the network contains four outputs, and the activation function is sigmoid (for the outputs to be between $0$ and $1$). The output of $l[\omega^l]$ is then renormalized via multiplication by $\overline{q}$ to obtain volumes between $0$ and $\overline{q}$. \\

To obtain optimal volumes of the market maker, we minimize the opposite of the Hamiltonian function defined by Equation \eqref{Hamiltonian Agent continuous version}. We generate $K>0$ random samples of $z$ and $q$, and conduct several epochs of batch learning with the following weights update:
\begin{align*}
    \omega^{l} \leftarrow \omega^{l} + \mu^l \frac{1}{K} \sum_{k=1}^K \nabla_{\omega^{l}} l[\omega^{l}](z_k, q_k) \Big(\nabla_{l}h^c(l[\omega^{l}](z_k, q_k), z_k, q_k) - \rho\big(&(q_k + l[\omega^{l}]^{b, l} + l[\omega^{l}]^{b, d} - \overline{q})_+ \\
    + &(q_k - l[\omega^{l}]^{a, l} - l[\omega^{l}]^{a, d} + \overline{q})_-\big)\Big),
\end{align*}
where $\mu^l$ is the learning rate. The term scaled by $\rho$ corresponds to a penalty employed to force quotes to stay in $\mathcal{A}$, so that $ l[\omega^{l}]^{i,l}+ l[\omega^{l}]^{i,d} \in [0,2\overline{q}], i \in \{a, b\}$. In our computations we use $\rho = 0.1$. \\

\begin{figure}[H]
    \centering
    \includegraphics[width=0.45\textwidth]{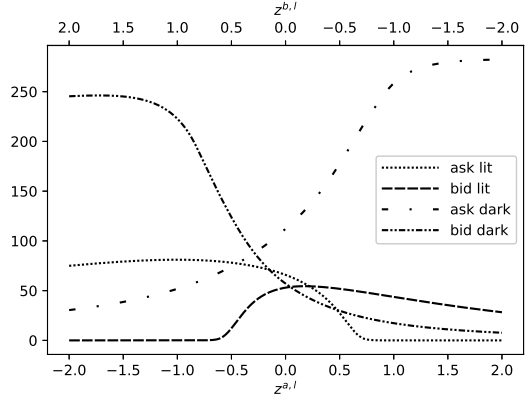}
    \caption{Best response of the market maker as a function of $z^{a,l}$ and $z^{b, l}$, with $q=50$.}
    \label{fig:l_zal_q0}
\end{figure}

Let us denote by $l^\star[\omega^l]$ the approximated optimal response function of the market maker $\mathcal{L}^\star$ (the result of the above optimization procedure). In Figure \ref{fig:l_zal_q0}, we see an example of the best response $l^\star[\omega^l]$ as a function of $z^{a,l} = -z^{b,l}$, when the market maker's inventory $q=50$ and other incentives $z^{a,d} = z^{b,d} = 0.05$ (close to zero). Remark that the choice of incentives is arbitrary only and aimed at reflecting the main properties of $l^\star[\omega^l]$.
The observed behavior has quite natural interpretation. The incentive $z^{a,l}$ is a remuneration of the market maker when his limit order is executed on the ask side of the lit pool. When this incentive increases, the market maker ensures to have a small imbalance on the ask side of the lit pool so that he can earn $z^{a,l}$. Because of his positive inventory, the volume posted on the ask side of the dark pool is higher than on the bid side of the dark pool: the market maker wants to liquidate his long position. Similarly when the incentive $z^{b,l}$ increases, the market maker wants to benefit from it when transacting on the bid side of the lit pool. This explains the small imbalance on the bid side of the lit pool for positive $z^{b,l}$. Mathematically speaking, the function $h^c$ is increasing in $z^{a,l}$. Thus for a high $z^{a,l}$, the value of the term $\mathcal{E}(z^{a,l}, l^{\star}[\omega^l]^{a,l})$ in the Hamiltonian is high. To benefit from the remuneration $z^{a,l}$, the intensity $\lambda^{a,l}$ must be high, which implies a small imbalance on the ask side, hence $I^a$ should be small. Similarly for $z^{b,l}$.\\

For $q=150$, $z^{b, l} = -z^{a, l}$, and other incentives $z^{a,d} = z^{b,d} = 0.05$ (close to zero), we display the volumes in Figure \ref{fig:l_zal_q50}:

\begin{figure}[H]
    \centering
    \includegraphics[width=8cm]{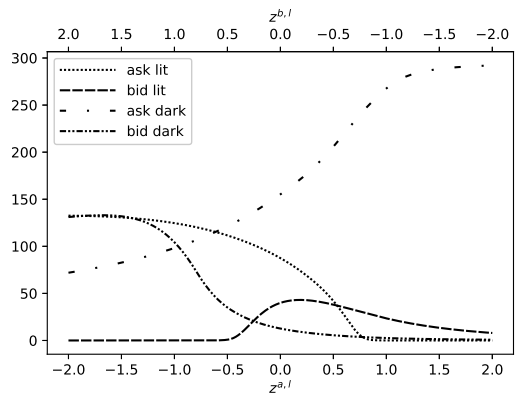}
    \caption{Best response of the market maker as a function of $z^{a,l}$ and $z^{b, l}$, with $q=150$.}
    \label{fig:l_zal_q50}
\end{figure}

As the market maker has a higher inventory, his quotes on the bid side of both pools decrease because of the inventory risk. Moreover, his quotes on the ask side of both pools increase to liquidate his long position. For high incentives $z^{b,l}$, a small volume on the bid side of the lit pool leads to a low imbalance on the bid side, hence a high probability of execution for passive ask orders in the dark pool, where the market maker tries to liquidate his position. Note that for high $z^{a,l}$, the imbalance is approximately equal to one half, because the market maker does not want to suffer from the latency effect (to be executed at the mid-price in the dark pool). \\

We now move to the problem of the principal. 

\subsubsection{An actor-critic approach to solve HJB equation \eqref{HJB Equation Principal First Form}}

A numerical approximation of the optimal incentives $z^\star$ can be obtained by 
\begin{enumerate}
    \item solving (numerically) the static maximization problem \eqref{Hamiltonian Agent continuous version}, which provides the approximation of the optimal response $\mathcal{L}^\star$ of the market maker,
    \item plugging this approximation in the continuous (with respect to $\mathcal{L}^\star$) version of Hamilton-Jacobi-Bellman equation $\eqref{HJB Equation Principal First Form}$, that is to say:
\end{enumerate}
\begin{align*}
\left\{
    \begin{array}{ll}
        0=\partial_{t}v(t,q) + \mathcal{H}^c\big(q,\mathcal{L}^\star,v(t,\cdot)\big), \quad q\in[-\overline{q},\overline{q}],t\in[0,T),  \\
        v(T,q)=-1,
    \end{array}
\right.
\end{align*}
with
\begin{align*}
\mathcal{H}^c\big(q,\mathcal{L}^\star,v(t,\cdot)\big):= \sup_{z\in \mathbb{R}^{5}}\mathcal{U}^c\big(z,q,\mathcal{L}^\star(z,q),v(t,\cdot)\big),
\end{align*}
and abusing the notation with $\mathcal{L}^\star$ denoting $\mathcal{L}^\star(z, q)$
\begin{align*}
& \mathcal{U}^c\big(z,q,\mathcal{L}^\star,v(t,\cdot)\big)  := v(t,q)\Big(\frac{\eta}{2}\sigma^{2}\gamma\big(z^{\tilde{S}}+q\big)^{2}+\frac{\eta^{2}\sigma^{2}}{2}\big(z^{\tilde{S}}\big)^{2}\Big)  \\ 
& + \sum_{\substack{i=\{a,b\} \\ j=\{l,d\}}} \lambda^{i,j}(\mathcal{L}^{\star l}) \bigg(\exp({\eta(z^{i,j}-c^{j}\ell^{\star i,j})})v\big(t,q-\phi(i) \ell^{\star i,j}\big)-v\big(t,q\big)\big(1+\eta\; \mathcal{E}(z^{i,j},\ell^{\star i,j})\big)\bigg).
\end{align*}
We obtain explicitly $z^{\tilde{S}} = -\frac{\gamma}{\gamma+\eta}q$, so we are only interested in finding optimal $(z^{a,l}, z^{b,l}, z^{a,d}, z^{b,d})$. The classical method to solve the above problem is to obtain an approximation of the value function \textit{via} a finite difference scheme on a grid. Since the size of the grid increases exponentially with the number of dimensions, using this approach is not possible for a high dimension. Therefore, to address our five-dimensional optimization problem, we resort to neural networks.\\

We use an algorithm known in reinforcement learning literature as the actor-critic method. The core of this approach is the representation of the value function and optimal controls with deep neural networks. The learning procedure itself consists of two stages: value function update (also called critic update) and controls update (actor update).  \\

We first split our problem into sub-problems corresponding to different time steps. We consider a time step $\Delta t$. The first-order approximation of the value function at time $t$ gives
\begin{align*}
   v(t, \cdot)\approx v(t+\Delta t, \cdot) - \partial_t v(t+\Delta t, \cdot).
\end{align*}

For each time step $\Delta t$, we represent the value function and the incentives with neural networks. Our procedure is backward in time, and we start from $T - \Delta t$, recalling that $v(T, \cdot) = -1$. Let us fix $t \in [0, T-\Delta t]$. Value function at time $t$ is represented by $v_t[\omega^{v_t}](\cdot)$ which is a feedforward neural network, parameterized by weights $\omega^{v_t}$, which approximates the value function corresponding to the current set of incentives approximated by the neural network $z_t[\omega^{z_t}](\cdot)$, parametrized by $\omega^{z_t}$. Critic's network is composed of $2$ hidden layers with $20$ nodes in each of these layers with ELU activation functions. The final layer of the network contains one output, and the activation is affine. Actor's network is composed of $2$ hidden layers with 20 nodes in each of these layers with ELU activation functions. The final layer of the network contains four outputs, and the activation is $\tanh$ (this allows the output to stay between $-1$ and 1), which is therefore renormalized by $\overline{z}$. The first step is the following update of the value function network's weights $\omega^{v_t}$:
\begin{align*}
    \omega^{v_t}\!\leftarrow \!\omega^{v_t} \!+ \!\mu^v\! \frac{1}{K}\! \sum_{k=1}^K\! \nabla_{\omega^{v_t}} \!v_t[\omega^{v_t}](q_k) \big(\!v_{t\!+\!\Delta t}[\omega^{v_{t\!+\!\Delta t}}](q_k)\! + \!\mathcal{U}^c(z_t[\omega^{z_t}](q_k), q_k, l^\star[\omega^l],\! v_{t\!+\!\Delta t}[\omega^{v_{t\!+\!\Delta t}}\!](\cdot)\!) \!-\! v_t[\omega^{v_t}]\!(q_k)\big)\!,
\end{align*}
where $\mu^v$ is a learning rate, $\mathcal{U}^c(z_t[\omega^{z_t}](q_k), q_k, l^\star[\omega^l], v_t[\omega^{v_t}](\cdot))$ corresponds to the function under the supremum of the Hamiltonian \eqref{hamiltonian} calculated using the current controls $z_t[\omega^{z_t}]$. The quantities $q_k, \; k\in\{1, \ldots, K\}$ are the elements of the training set, more precisely $K$ uniformly distributed elements from the interval~$[-\overline{q}, \overline{q}]$. We use $\mathcal{U}(z_t[\omega^{z_t}](q_k), q_k, l^\star[\omega^l], v_{t+\Delta t}[\omega^{v_{t+\Delta t}}](\cdot))$ as an approximation of $\partial_t v(t+\Delta t, \cdot)$ to apply the first order approximation of the value function described above. \\
 
When the value function's neural network approximates the value function corresponding to the current control $z_t[\omega^{z_t}]$, we can move to the stage of optimization over control values (also called policy update in reinforcement learning literature). Our policy update consists of two different procedures. The first one is an exploitation phase where the weights are updated according to the best direction suggested by the gradient of the function  $\mathcal{U}^c(z_t[\omega^{z_t}](q_k), q_k, l^\star[\omega^l], v_t[\omega^{v_t}](\cdot))$:
\begin{align*}
    \omega^{z_t} \leftarrow \omega^{z_t} + \mu^z \frac{1}{K} \sum_{k=1}^K \nabla_{\omega^{z_t}} {z_t}[\omega^{z_t}](q_k) \nabla_{z_t}\mathcal{U}^c\big(z_t[\omega^{z_t}](q_k), q_k, l^\star[\omega^l], v_t[\omega^{v_t}](\cdot)\big),
\end{align*}
where $\mu^z$ is a learning rate. This type of updates is usually called policy gradient.  \\

Another type of updates we use in the learning procedure is an exploration phase. During this phase, we use the current values given by the neural network of controls and introduce noise to these values, to explore the values slightly different from those proposed by the neural network. Noise is normally distributed around 0 with standard deviation chosen beforehand (in the following examples, we use standard normal distribution). This phase could help us to quit local minima, in case the algorithm is trapped in one. The following updates characterize this phase:
\begin{align*}
    \omega^{z_t} \leftarrow \omega^{z_t} + \hat{\mu}^z \frac{1}{K} \sum_{k=1}^K \varepsilon_k \nabla_{\omega^{z_t}} {z_t}[\omega^{z_t}](q_k) \Big(&\mathcal{U}^c\big(z_t[\omega^{z_t}](q_k) + \varepsilon, q_k, l^\star[\omega^l], v_t[\omega^{v_t}](\cdot)\big) \\
    & - \mathcal{U}^c\big(z_t[\omega^{z_t}](q_k), q_k, l^\star[\omega^l], v_t[\omega^{v_t}](\cdot)\big)\Big),
\end{align*}
where $\varepsilon$ is a vector of length $K$ representing introduced perturbations and $\hat{\mu}^z$ is a learning rate.

\subsection{Numerical Results}\label{section numerical results}
In the following we consider $\Delta t = 1$. Since time has little impact on the quotes chosen by the market maker (see \cite{el2018optimal, gueant2013dealing}), we present the results only for time $T-1$, the extension to earlier time steps is straightforward. As mentioned before, the optimization problems considered are symmetric with respect to the inventory variable $q$. 

\subsubsection{Reference model without the exchange}
First, we present a reference model without the intervention of the exchange. We consider the following parameters:
\begin{itemize}[itemsep=0.3pt]
    \item Risk aversion of the market maker: $\gamma=0.01$;
    \item Market impacts: $\Gamma^l=10^{-4},\Gamma^d=5\times 10^{-5}$;
    \item Influence of the imbalance on the orders arrival: $\theta^l=\theta^d=0.15$;
    \item Volatility: $\sigma=0.1$;
    \item Fees: $c^l=0.05, c^d=0.01$;
    \item Order flow intensity parameter: $A^l=5\times 10^{3}, A^d=3\times 10^3$.
\end{itemize}

\begin{figure}[H]
\begin{center}
    \includegraphics[width=0.45\textwidth]{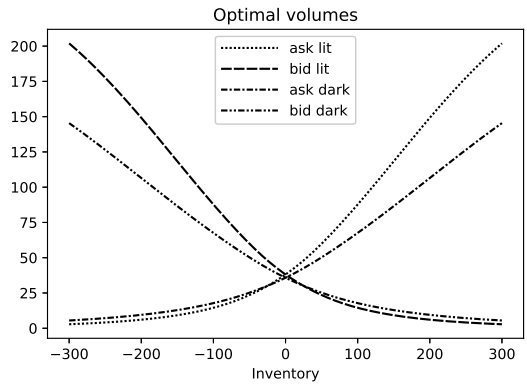}
    \caption{Optimal quotes of the market maker.}\label{benchmark no exchange}
\end{center}
\end{figure}

In Figure \ref{benchmark no exchange}, we present the optimal quotes of the market maker. 

One can see that the market maker splits his orders equitably between the lit and dark pools when his inventory is near zero. However, when he has a very positive (resp. negative) inventory, he has a large imbalance on the ask (resp. bid) side of the lit pool, to liquidate his position in the dark pool. Such behavior shows that the market maker uses the dark pool as a way to liquidate a large position by adjusting the imbalance in the lit pool. Indeed, when he posts a high volume on the ask side of the lit pool, he encourages ask orders in the dark pool. Thus, as he prioritizes the execution of a large ask order, he accepts to be executed at the mid-price in the dark pool. When $q=300$, he does not post a sell order of size $300$ in the dark pool, because of the quadratic variation between the mid-price and its inventory process (which can be seen as a quadratic penalty in the market maker's PnL process with respect to the volumes displayed). Because of the latency generated on the ask side of the lit pool, the market takers sending market orders on the bid side of the dark pool are likely to be executed at an unfavorable price. This is why the market maker posts a non-zero volume on the bid side of the dark pool. Remark also that for small inventories, the market maker posts volumes on both ask and bid sides of the dark pool because he may accept to increase his inventory risk by being executed at a more favorable price in the dark pool due to the latency effect (the volumes displayed in the lit pool lead to 50 percents chance to face this effect at least on one of the sides of the dark pool). Note that the parameters $A^l>A^d$ describe the fact that there are, on average, much more orders in the lit pool than in the dark pool.\footnote{This assumption is consistent with the MIFID II regulation rolled out on January 3, 2018, which imposes a cap on volumes traded in the dark pools.} \\

In the following sections, we present several numerical experiments involving the incentive policy of the exchange.

\subsubsection{Reference model with the exchange}
In this section, we present a reference model with the exchange. We take the same parameters as in the case without the exchange, and we set the exchange's risk aversion: $\eta=0.02$.\\

\begin{figure}[H]

\begin{minipage}[c]{.46\linewidth}
    \begin{center}
            \includegraphics[width=0.97\textwidth]{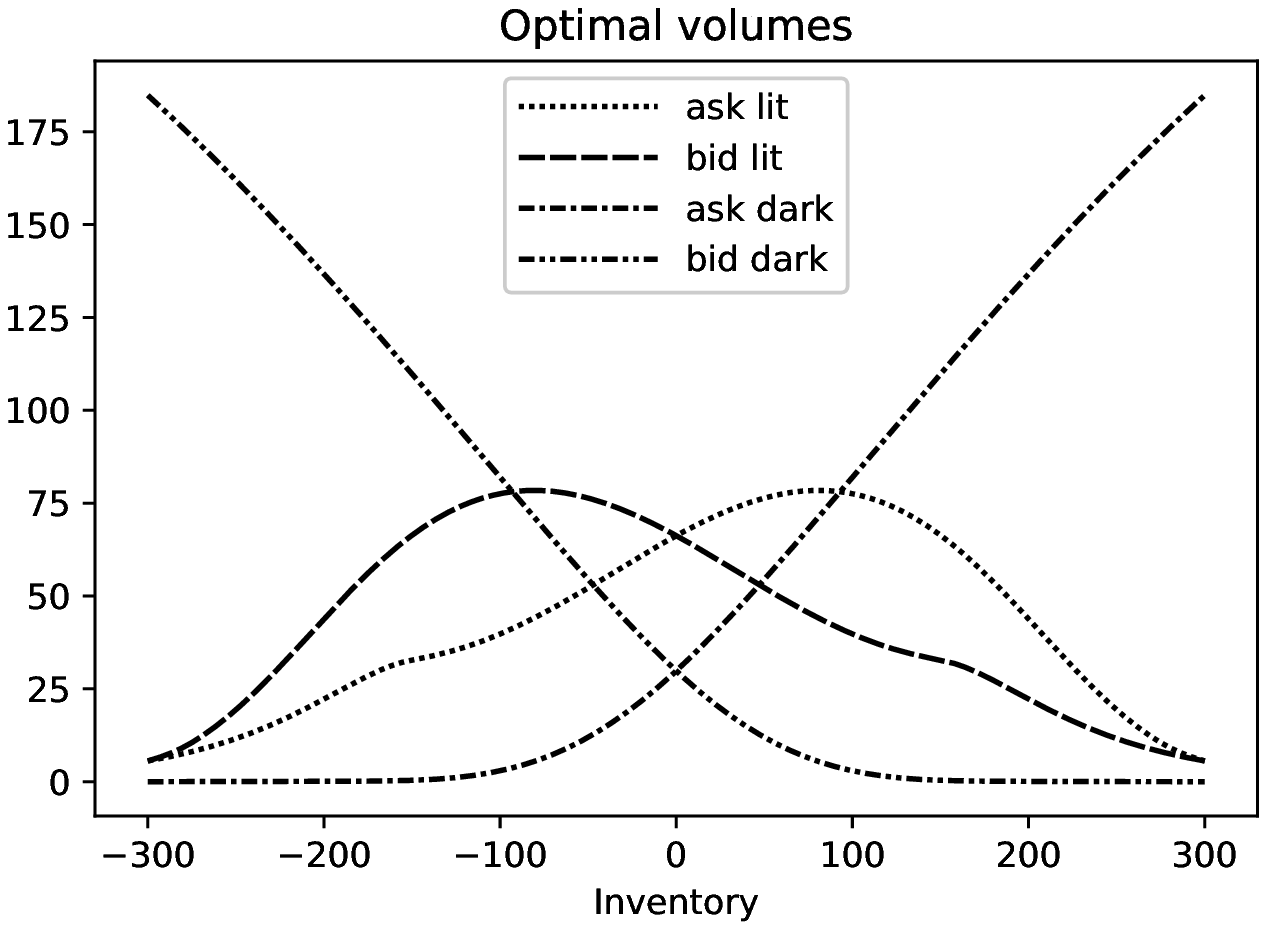}
            \vspace{-3mm}
            \caption{Optimal quotes of the market maker.}\label{Optimal quotes bench}
        \end{center}
\end{minipage} \hfill
\begin{minipage}[c]{.46\linewidth}
     \begin{center}
             \includegraphics[width=0.97\textwidth]{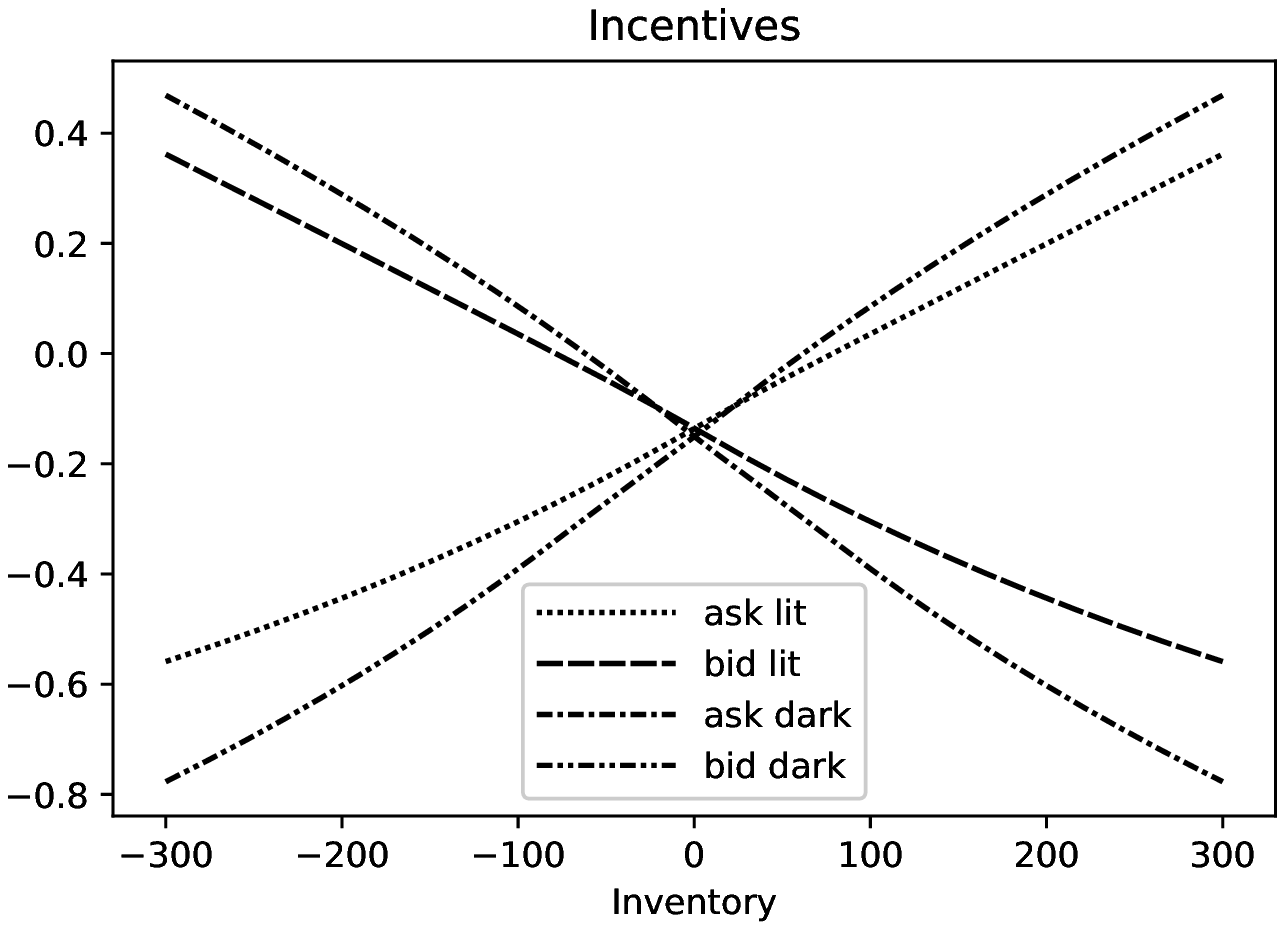}
             \vspace{-3mm}
             \caption{Optimal incentives of the exchange.}\label{Optimal incentives bench}
         \end{center}
   \end{minipage} 
\end{figure}

In Figures \ref{Optimal quotes bench} and \ref{Optimal incentives bench}, we present the optimal quotes of the market maker and the optimal incentives provided by the exchange. 
\vspace{-3mm}

\vspace{-3mm}
The presence of incentives has significant effects on the market maker's behavior. When the market maker has an inventory near zero, incentives lead to an increase of the volumes posted in the lit pool and a decrease of that in the dark pool compared to Figure \ref{benchmark no exchange}. Thus the exchange improves the liquidity in the lit venue. Moreover, the strategy of the market maker for very positive or negative inventory is modified. When he has a very positive inventory, he posts a higher volume on the ask side of the dark pool than in the case without exchange. In addition to this, he posts an equal volumes (small but not negligible) on the ask and bid sides of the lit pool. So we see that the exchange prevents the market maker from artificial manipulation of the market, consisting in creation of high imbalance on the ask side. As the imbalance is around $1/2$, the market maker does not take advantage of the latency effect. \\

In Figure \ref{Optimal incentives bench}, we see that, even if our problem is much more intricate than those of \cite{baldacci2019optimal, el2018optimal}, the shape of the principal's incentives are essentially linear functions of the market maker's inventory. 

\subsubsection{High volatility regime}
We now investigate the impact of higher volatility on the posted volumes with and without the exchange. We take $\sigma=0.4$, the other parameters being as previously. 
\vspace{-4mm}
\begin{figure}[H]
\begin{minipage}[c]{.46\linewidth}
    \begin{center}
            \includegraphics[width=0.97\textwidth]{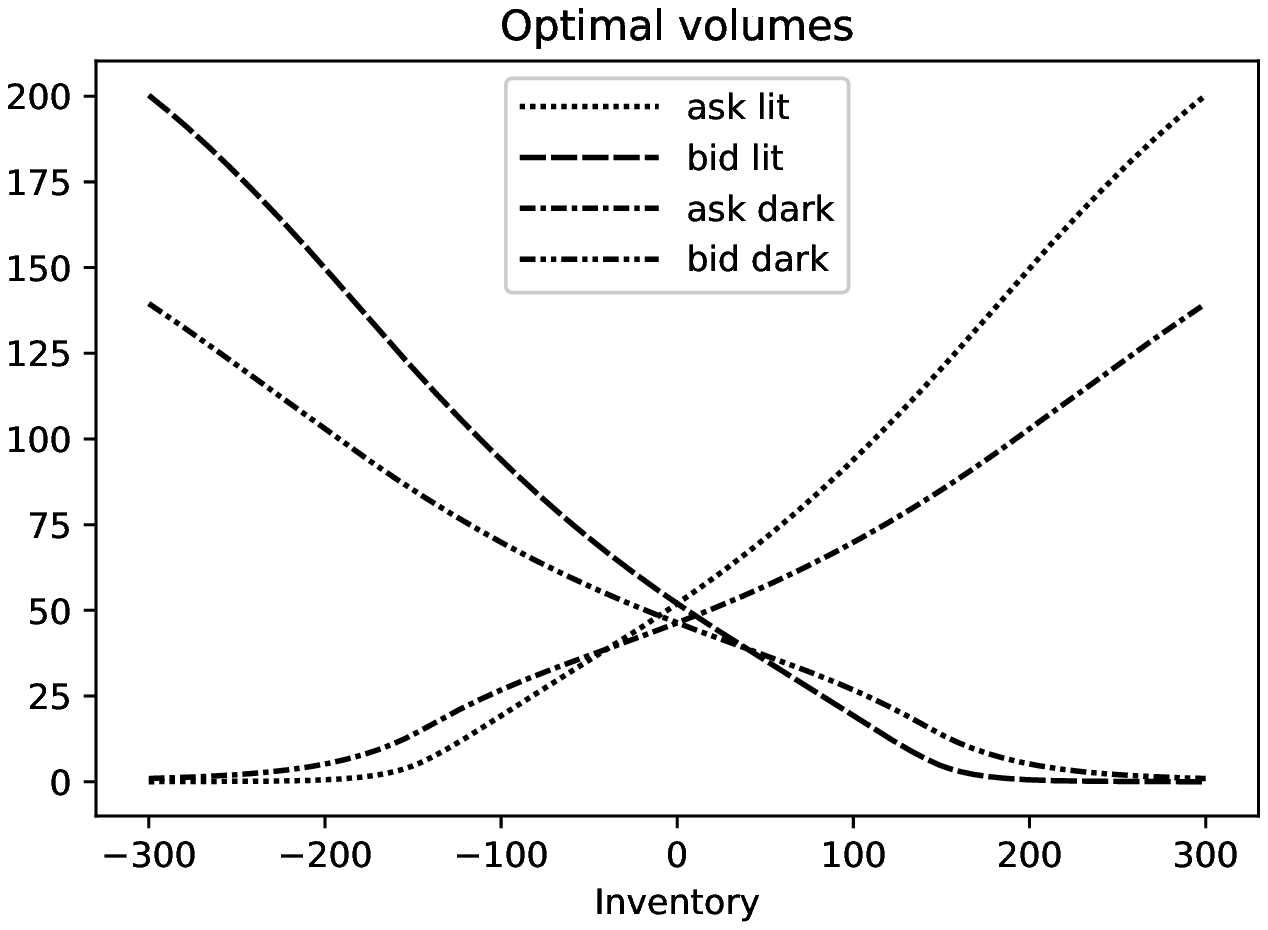}
            \vspace{-3mm}
            \caption{Optimal quotes of the market maker without the exchange.}\label{Optimal incentives higher vol}
        \end{center}
\end{minipage} \hfill
\begin{minipage}[c]{.46\linewidth}
     \begin{center}
             \includegraphics[width=0.97\textwidth]{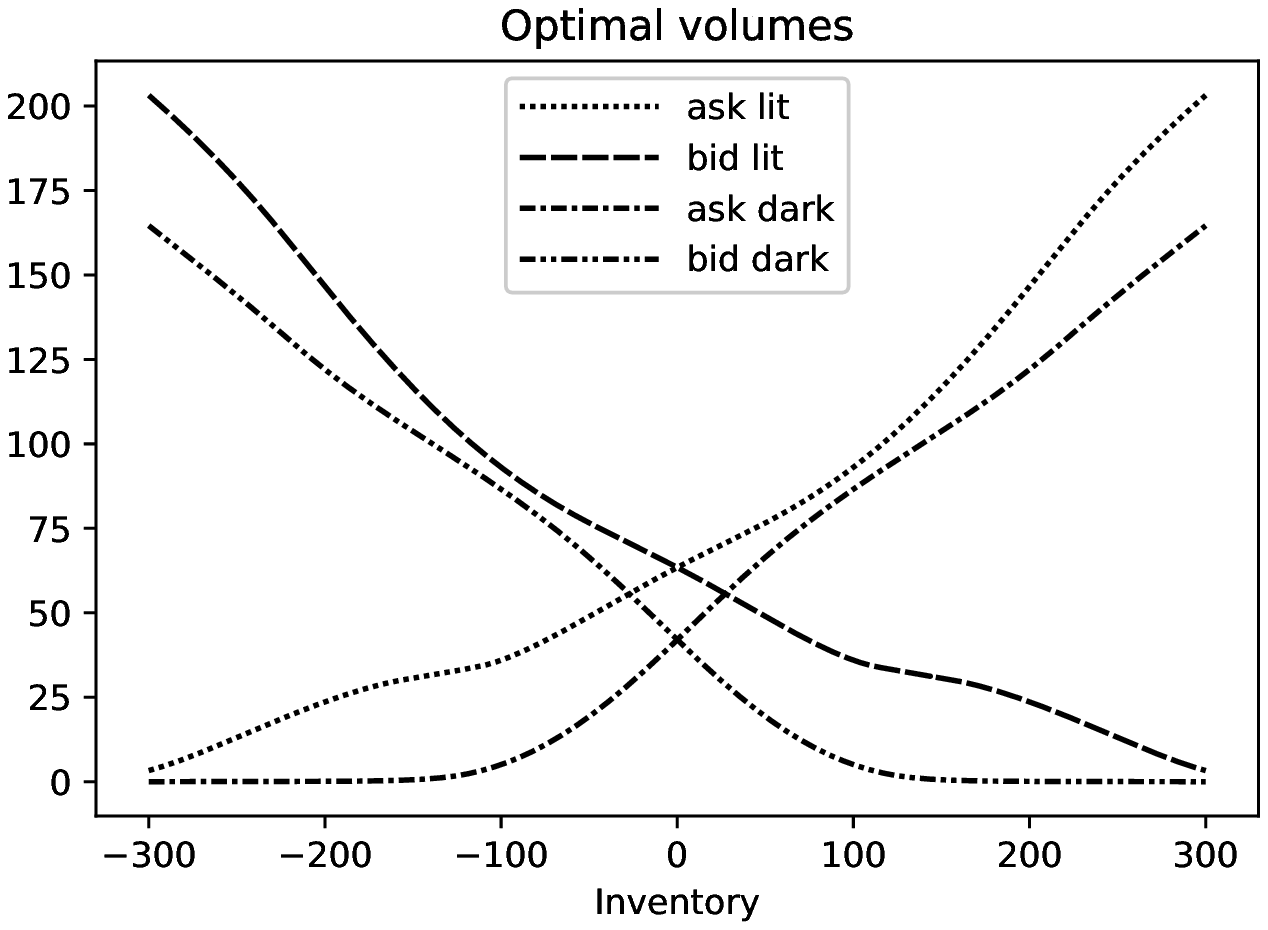}
             \vspace{-3mm}
             \caption{Optimal quotes of the market maker with the exchange.}\label{Optimal quotes higher vol}
         \end{center}
   \end{minipage} 
  
\end{figure}

In Figure \ref{Optimal incentives higher vol} we see that, compared to Figure \ref{benchmark no exchange}, higher volatility does not change significantly the strategy of the market maker without the exchange. We observe that the contract has more limited influence in the case of high volatility as the market maker follows the same strategy as without exchange. In particular, he does not keep his imbalance equal to $1/2$ when he has a very positive or negative inventory. This is because higher volatility leads to an increase in market activity, and the market maker is more willing to send higher volumes on the side of interest of both pools. 

\subsubsection{Same parameters for the lit and dark pools}
Here we show the volumes posted by the market maker and the incentives of the exchange, when the lit and dark pools share the same characteristics. We consider the following set of parameters: 
\vspace{-1mm}
\begin{itemize}[itemsep=0.3pt]
    \item Risk aversion of the market maker and of the exchange respectively: $\gamma=0.01,\eta=0.02$;
    \item Market impacts: $\Gamma^l=\Gamma^d=10^{-4}$;
    \item Influence of the imbalance on the orders arrival: $\theta^l=\theta^d=0.2$;
    \item Volatility: $\sigma=0.2$;
    \item Fees: $c^l=c^d=0.05$;
    \item Order flow intensity parameters: $A^l=A^d=5\times 10^3$.
\end{itemize}

\begin{figure}[H]
   \begin{minipage}[c]{.46\linewidth}
    \begin{center}
            \includegraphics[width=0.97\textwidth]{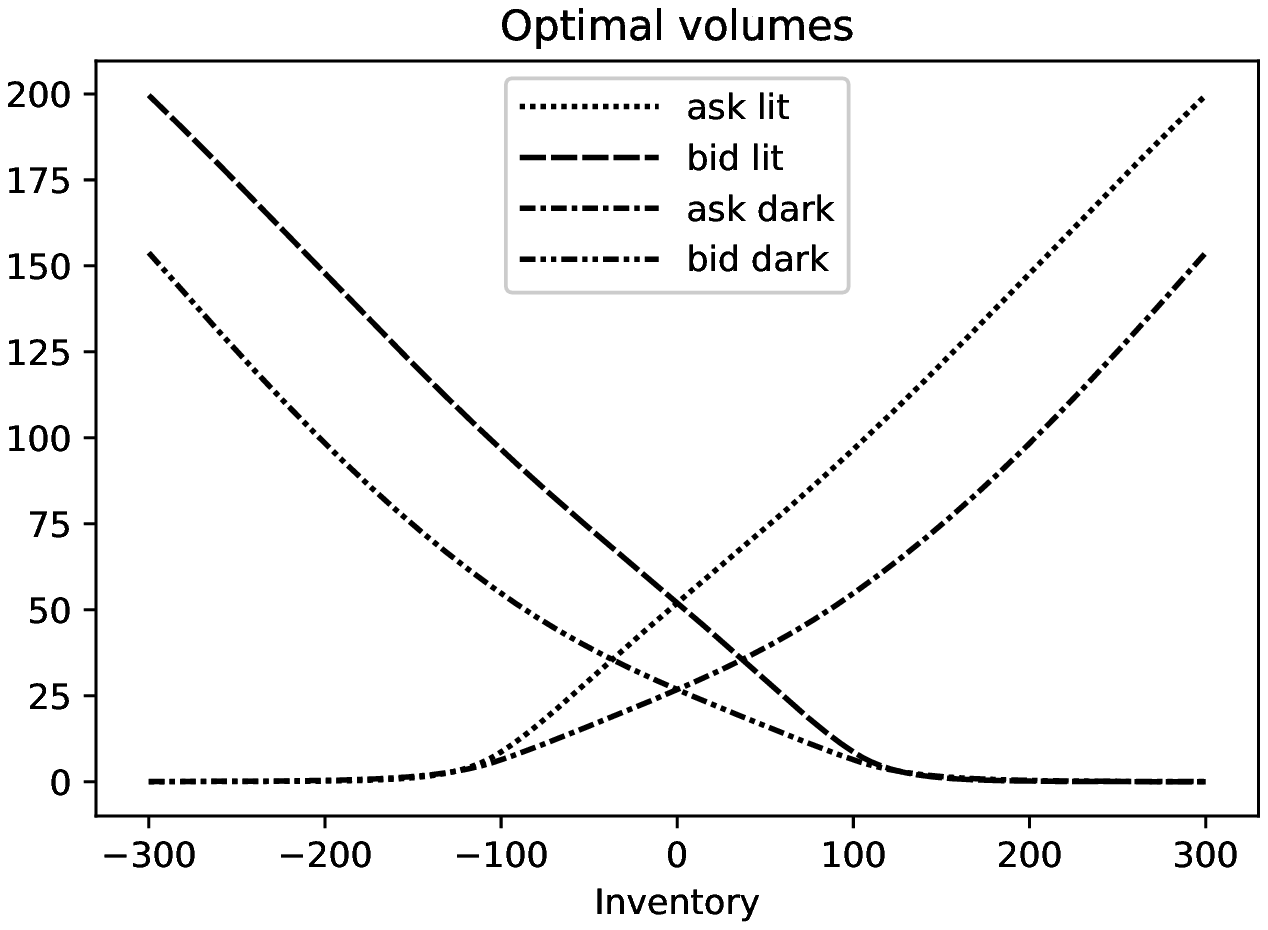}
            \vspace{-3mm}
            \caption{Optimal quotes of the market maker without the exchange.}\label{Optimal incentives less}
        \end{center}
\end{minipage} \hfill
\begin{minipage}[c]{.46\linewidth}
     \begin{center}
           \includegraphics[width=0.97\textwidth]{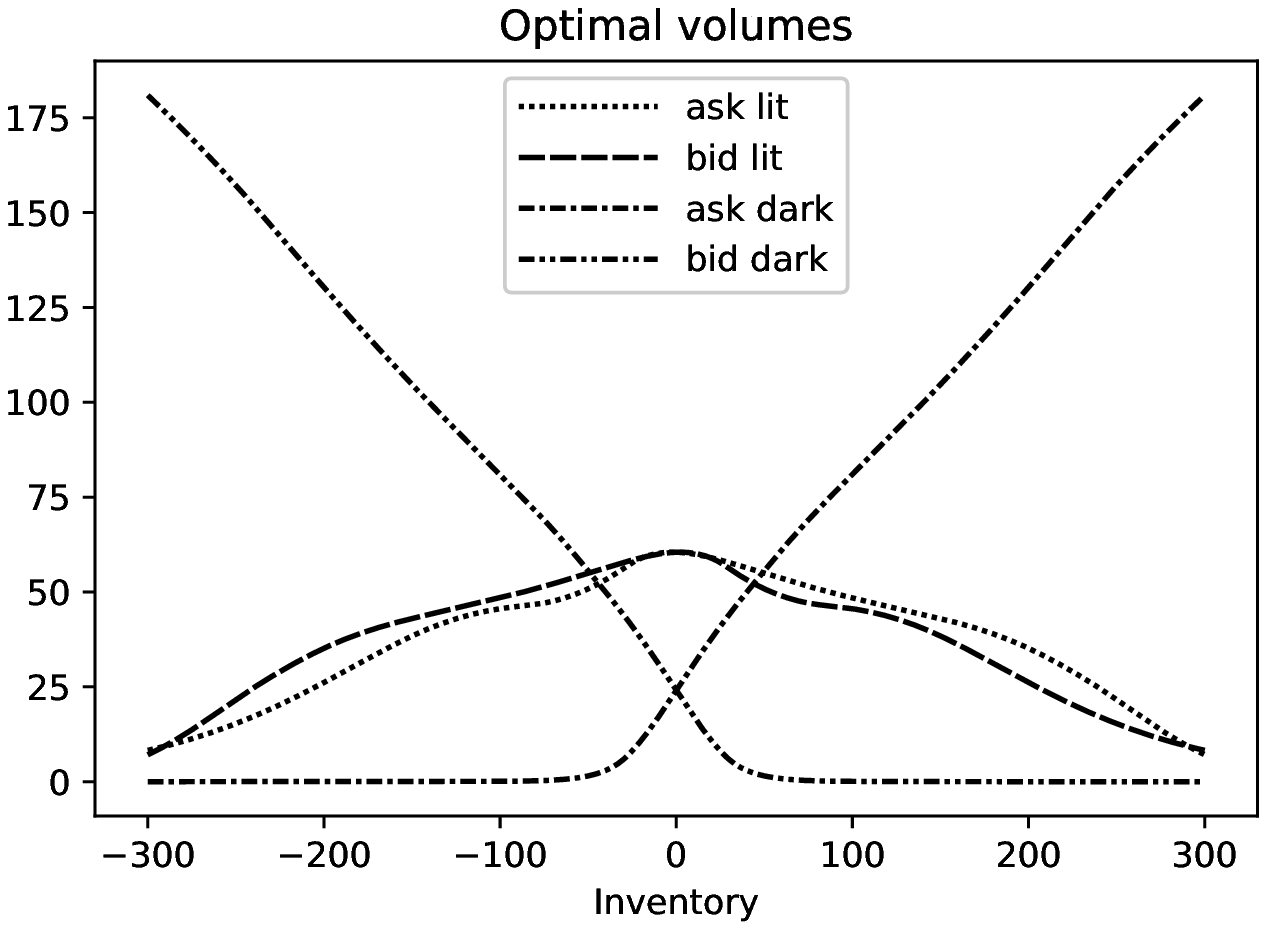}
             \vspace{-3mm}
             \caption{Optimal quotes of the market maker with the exchange.}\label{Optimal quotes less}
         \end{center}
   \end{minipage}
\end{figure}

In Figures \ref{Optimal incentives less} and \ref{Optimal quotes less}, we see that the repartition of volumes between the lit and dark pools has not changed significantly compared to the reference case with and without contract. The main difference is that, in the absence of the exchange, the market maker posts higher volumes in the lit pool compared to the dark one when he has a small inventory. It happens because the dark pool does not provide lower market impact and transaction costs contrary to the reference case. Keeping his imbalance near $1/2$ for small inventories, the market maker still does not take advantage of the latency effect. Finally, in both cases, the dark pool is still used by the market maker as a way to liquidate a high inventory.  

\vspace{-3mm}

\subsubsection{High market impact in the lit pool}
We now show the volumes displayed by the market maker with and without the exchange, when the parameters of the dark pool make it more appealing than the lit pool. In particular, the market impact in the dark pool is five times smaller than in the lit pool. We consider the following set of parameters: 

\begin{itemize}[itemsep=0.3pt]
    \item Risk aversion of the market maker and of the exchange respectively: $\gamma=0.01,\eta=0.02$;
    \item Market impacts: $\Gamma^l=10^{-4}, \Gamma^d=2\times 10^{-5}$;
    \item Influence of the imbalance on the orders arrival: $\theta^l=\theta^d=0.15$;
    \item Volatility: $\sigma=0.1$;
    \item Fees: $c^l=0.05,c^d=0.01$;
    \item Order flow intensity parameters: $A^l=5\times 10^{3}, A^d=3\times 10^3$.
\end{itemize}

\begin{figure}[H]
   \begin{minipage}[c]{.46\linewidth}
    \begin{center}
            \includegraphics[width=0.97\textwidth]{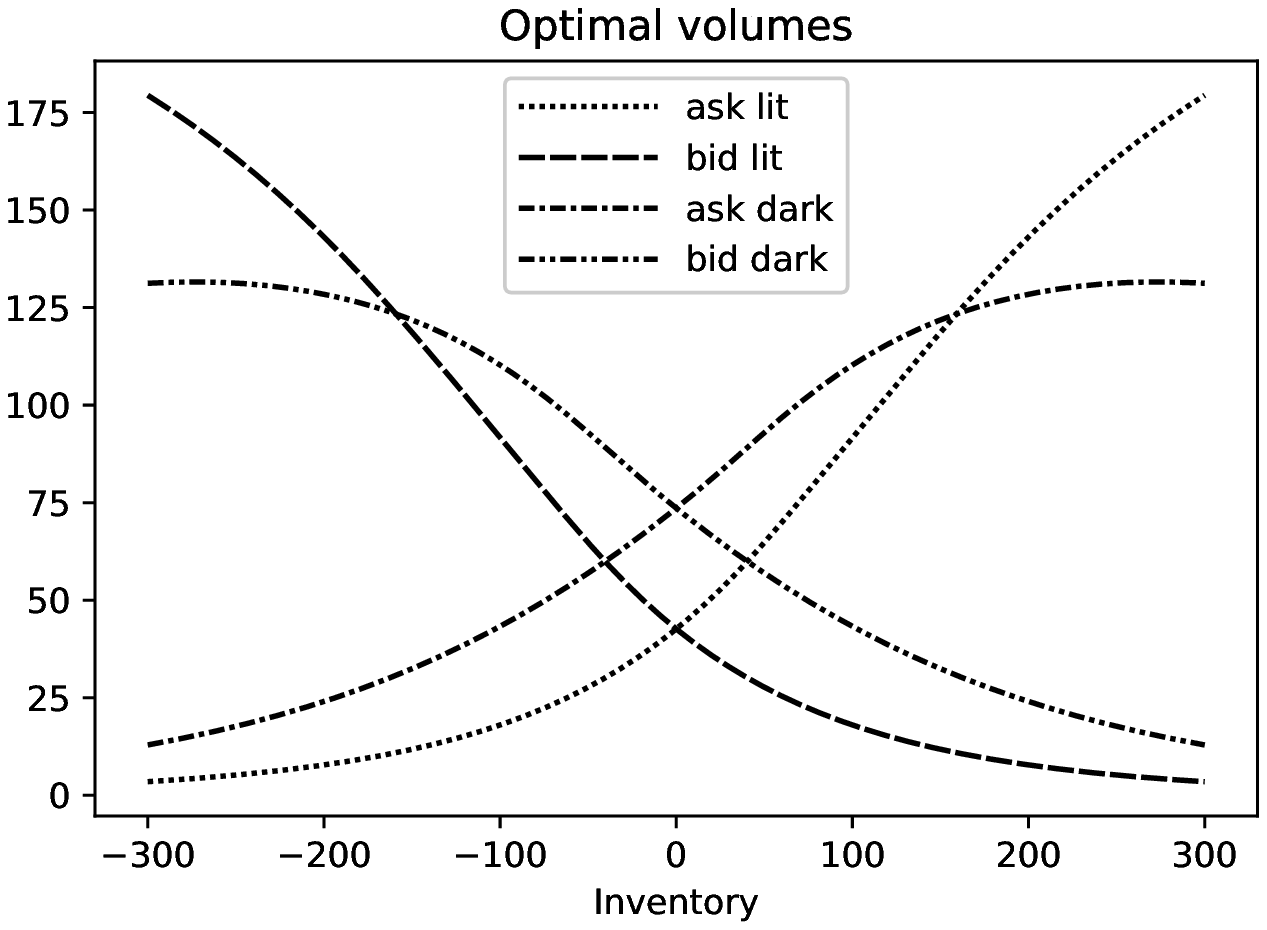}
            \caption{Optimal quotes of the market maker without the exchange.}\label{Optimal incentives more}
        \end{center}
\end{minipage} \hfill
\begin{minipage}[c]{.46\linewidth}
     \begin{center}
     \includegraphics[width=0.97\textwidth]{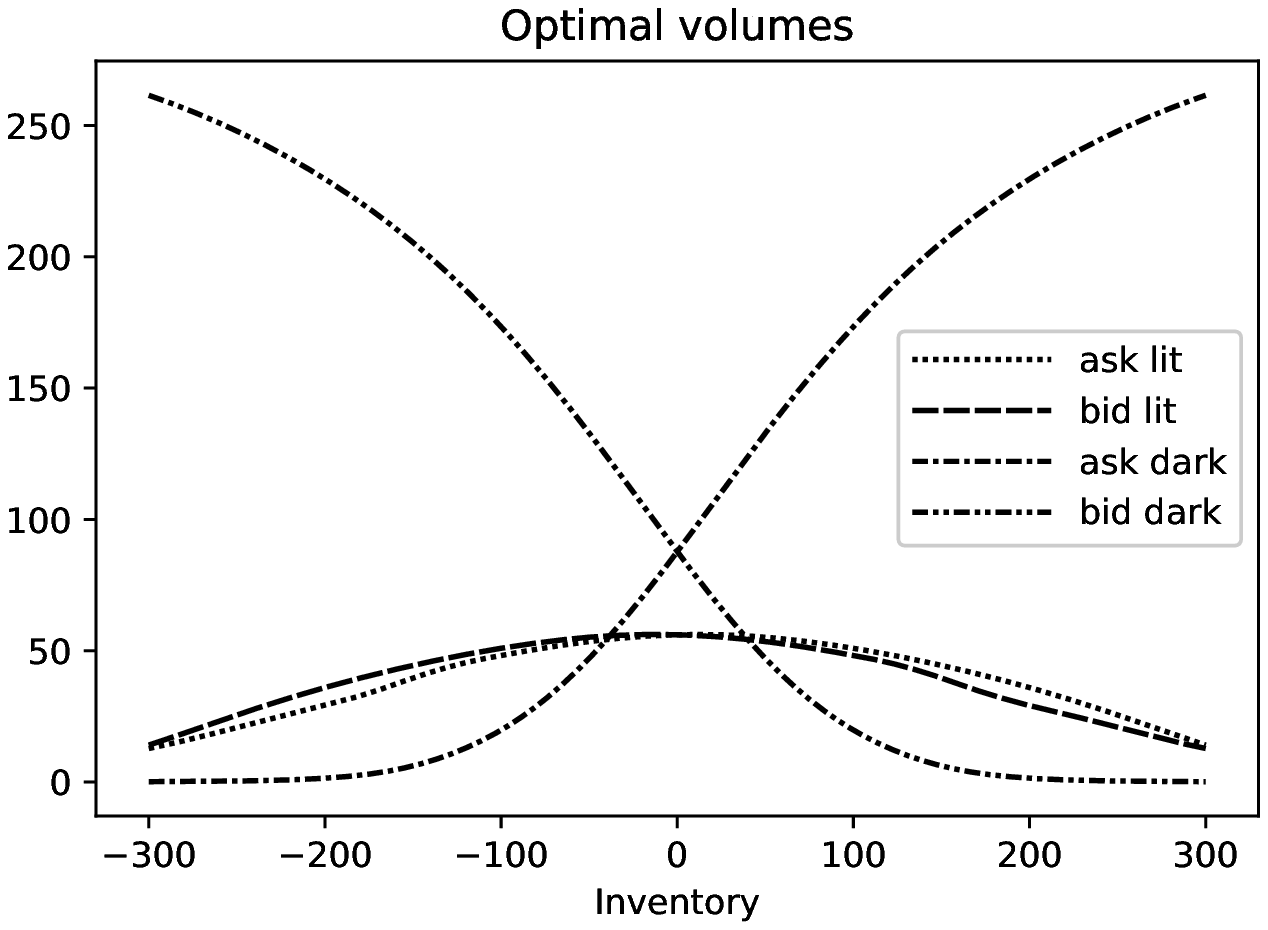}
             \caption{Optimal quotes of the market maker with the exchange.}\label{Optimal quotes more}
         \end{center}
   \end{minipage} 
\end{figure}

In Figures \ref{Optimal incentives more} and \ref{Optimal quotes more}, we see the influence of a higher market impact and transaction costs in the lit pool. Either with or without the intervention of the exchange and for small inventories, the market maker posts higher volumes in the dark pool than in the lit pool. We recover similar behavior for the displayed volumes as in the reference case with and without the exchange in Figures \ref{benchmark no exchange} and \ref{Optimal quotes bench}.

\subsubsection{High market impact on both venues}
In this last section, we show how the volumes are split between the lit and dark pools when the market impact in the lit and dark pools are equal. We consider the following set of parameters: 
\begin{itemize}[itemsep=0.3pt]
    \item Risk aversion of the market maker and of the exchange respectively: $\gamma=0.01,\eta=0.02$;
    \item Market impacts: $\Gamma^l=\Gamma^d=2.5\times 10^{-4}$;
    \item Influence of the imbalance on the orders arrival: $\theta^l=\theta^d=0.15$;
    \item Volatility: $\sigma=0.1$;
    \item Fees: $c^l=0.05,c^d=0.01$;
    \item Order flow intensity parameters: $A^l=5\times 10^{3}, A^d=3\times 10^3$.
\end{itemize}

\begin{figure}[H]   
\begin{minipage}[c]{.46\linewidth}
    \begin{center}
            \includegraphics[width=0.97\textwidth]{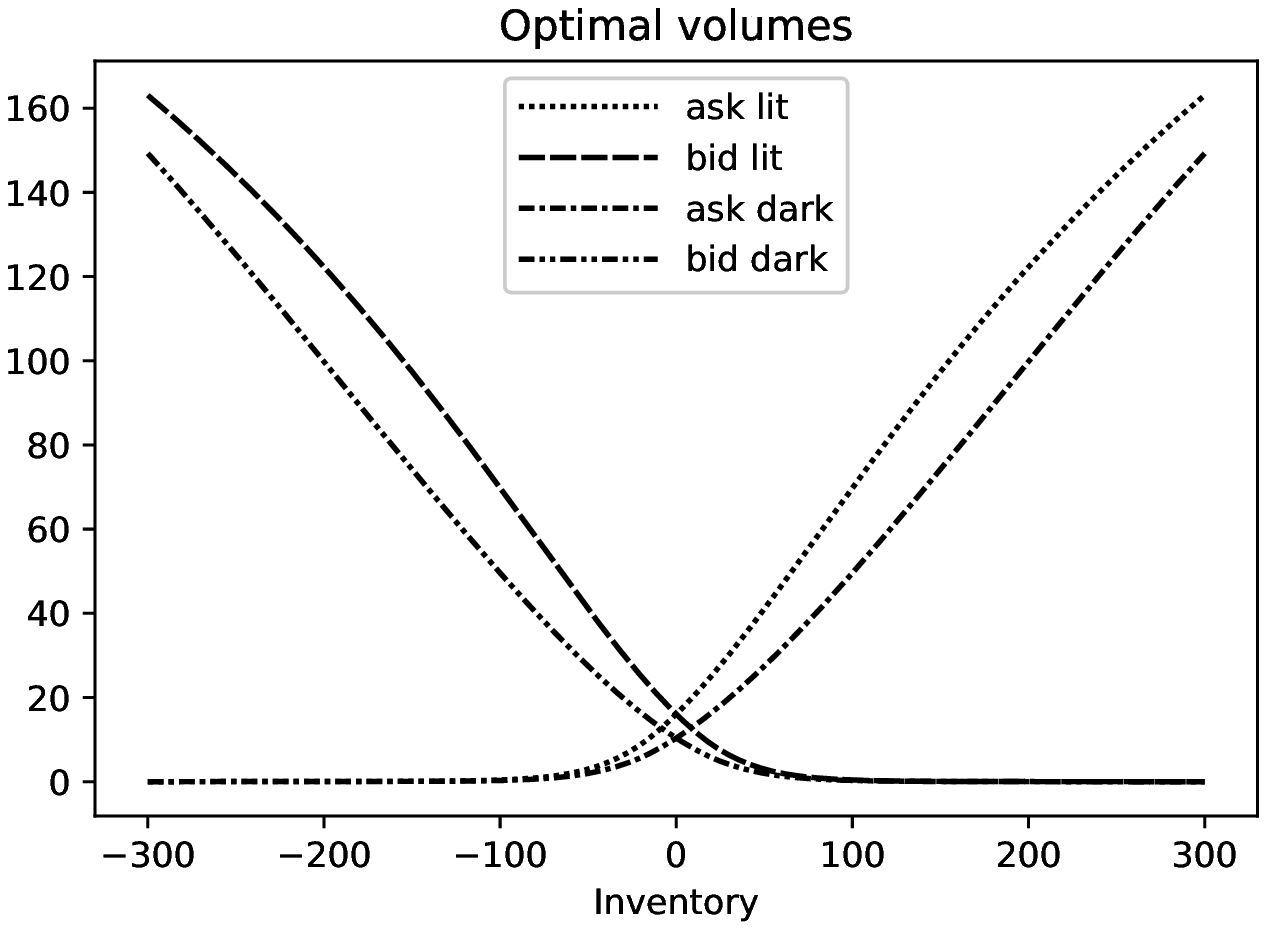}
            \caption{Optimal quotes of the market maker without the exchange.}\label{Optimal incentives MI}
        \end{center}
\end{minipage} \hfill
\begin{minipage}[c]{.46\linewidth}
     \begin{center}
             \includegraphics[width=0.97\textwidth]{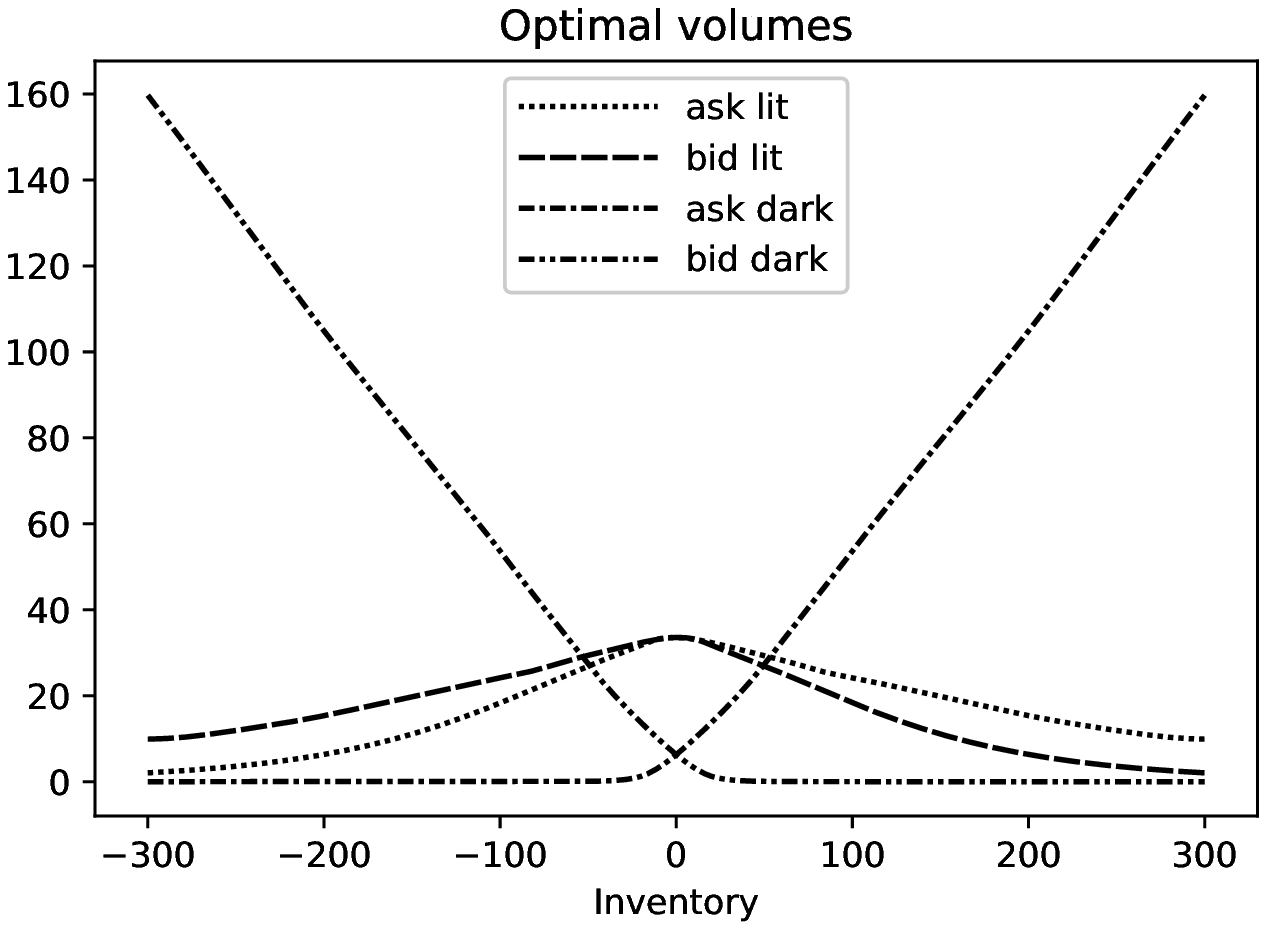}
             \caption{Optimal quotes of the market maker with the exchange.}\label{Optimal quotes MI}
         \end{center}
   \end{minipage} 
\end{figure}

In Figures \ref{Optimal incentives MI} and \ref{Optimal quotes MI}, we see that a higher market impact reduces the volume posted on both lit and dark pools. We also recover a behavior similar to the reference case without the exchange. For the case with the exchange, for very positive (resp. negative) inventory, the market maker has an ask (resp. bid) imbalance slightly above $1/2$, meaning that market takers on the bid (resp. ask) of the dark pool are more likely to be executed at a price unfavorable for them due to the latency effect. 

\begingroup
\setcounter{section}{0}
\renewcommand\thesection{\Alph{section}}
\section{Appendix}

\subsection{Dynamic programming principle and contract representation}\label{Contract representation}

For any $\mathbb{F}$ stopping time $\tau \in [t,T]$ and $\mathcal{L} \in \mathcal{A}_{\tau}$, we define: 
\begin{align*}
&J_{T}(\tau,\mathcal{L})=\mathbb{E}_{\tau}^{\mathcal{L}}\bigg[-\mathcal{D}_{\tau,T}(\mathcal{L})\text{exp}\big(-\gamma\xi\big)\bigg], \quad V_\tau=\sup_{\mathcal{L}\in\mathcal{A}_\tau}J_{T}(\tau,\mathcal{L}),
\end{align*}
where $\mathcal{A}_{\tau}$ denotes the restriction of $\mathcal{A}$ to controls on $[\tau,T]$ and 
\begin{align*}
\mathcal{D}_{\tau,T}(\mathcal{L}):=\text{exp}\bigg(-\gamma\Big(\int_{\tau}^{T} \sum_{i\in \{a,b\}}\Big(\frac{\mathcal{T }}{2}\ell_{t}^{i,l}\mathrm{d}N_{t}^{i,l} +\sum_{\kappa\in K}\phi^{lat}(\kappa)\frac{\mathcal{T}}{2}\ell_{t}^{i,d}\mathrm{d}N_{t}^{i,d,\kappa}\Big)+Q_{t}\mathrm{d}S_{t}+\mathrm{d} \big[ Q_{\cdot},S_{\cdot} \big]_t\Big)\bigg),
\end{align*}
where 
\begin{align*}
\mathrm{d} \big[ Q_{\cdot},S_{\cdot} \big]_t = -\sum_{\substack{i\in\{a,b\}\\ j\in\{l,d\}}}\sum_{k\in\mathcal{V}^j}\Gamma^j k^2 \mathrm{d}N_t^{i,j,k}.    
\end{align*}
We now set the dynamic programming principle associated to the control problem \eqref{Market Maker's Problem with exchange}.

\begin{lemma}\label{Lemma 6.2}
Let $t\in [0,T]$ and $\tau$ be an $\mathbb{F}$ stopping time with values in $[t,T]$. Then
\begin{align*}
&V_t=\underset{\mathcal{L}\in \mathcal{A}}{\textup{ess sup }}\mathbb{E}_{t}^{\mathcal{L}}\bigg[-\mathcal{D}_{t,\tau}(\mathcal{L})V_{\tau}\bigg].  
\end{align*}
\end{lemma}
The proof can be found in \cite[Lemma A.4]{el2018optimal}.

\subsection{Proof of Theorem \ref{Theorem Market Maker}}
\label{section:preuveagent}
To prove that $\mathcal{C} =\Xi$, we proceed in six steps. Our approach is largely inspired by \cite{el2018optimal}. However, for the sake of completeness, we provide here the details. \\

\textbf{Step 1:} For $\mathcal{L} \in \mathcal{A}$ it follows from the dynamic programming principle of Lemma \ref{Lemma 6.2} that the process 
\begin{align*}
& U_{t}^{\mathcal{L}}=V_{t}\mathcal{D}_{0,t}(\mathcal{L})
\end{align*}

defines a $\mathbb{P}^{\mathcal{L}}$-supermartingale for any $\mathcal{L}\in \mathcal{A}$. By standard analysis, we may then consider it  in  its  càdlàg  version  (by  taking  right  limits  along  rationals). By  the  Doob-Meyer decomposition, we can write
$U_{t}^{\mathcal{L}}=M_{t}^{\mathcal{L}}-A_{t}^{\mathcal{L}}$ where $M^{\mathcal{L}}$ is a $\mathbb{P}^{\mathcal{L}}$-martingale and $A_t^{\mathcal{L}}=A_{t}^{\mathcal{L},c}+A_{t}^{\mathcal{L},d}$ is an integrable non-decreasing predictable process such that $A_{0}^{\mathcal{L},c}=A_{0}^{\mathcal{L},d}=0$ with pathwise continuous component $A^{\mathcal{L},c}$ and with $A^{\mathcal{L},d}$ a piecewise constant predictable process. \\

From the martingale representation theorem under $\mathbb{P}^{\mathcal{L}}$, see Appendix A.1 in \cite{el2018optimal}, there exists $\tilde{Z}^{\mathcal{L}}\!\!=\!\!(\tilde{Z}^{\mathcal{L},S},\tilde{Z}^{\mathcal{L},i,j,k})_{i\in \{a,b\}, j\in \{l,d\},k\in\mathcal{V}^j}$  predictable, such that

\begin{align*}
M_{t}^{\mathcal{L}}=V_{0}+\int_{0}^{t}\tilde{Z}_{r}^{\mathcal{L},S}\mathrm{d}\tilde{S}_{r}+\sum_{\substack{i\in \{a,b\}\\j\in\{l,d\}}}\sum_{k\in\mathcal{V}^j}\int_{0}^{t}\tilde{Z}_{r}^{\mathcal{L},i,j}\mathrm{d}N_{r}^{\mathcal{L},i,j}.
\end{align*}

\textbf{Step 2:} We now show that $V$ is a negative process.  Thanks to the uniform boundedness of $\mathcal{L}\in \mathcal{A}$ and $I^{a},I^{b}\in [0,1]$ we get that 
\begin{align*}
\frac{L_{T}^{\mathcal{L}}}{L_{t}^{\mathcal{L}}}\geq \alpha_{t,T}= \text{exp}\bigg(\!\!-\!\!\sum_{j\in \{l,d\}}\frac{\theta^{j}}{\sigma}(N_{T}^{a,j}-N_{t}^{a,j}+N_{T}^{b,j}-N_{t}^{b,j})-2(A^j-\epsilon)(T-t)\bigg).    
\end{align*}
Therefore using the definition of $\mathcal{D}_{t,T}(\mathcal{L})$, we obtain
\begin{align*}
V_{t}\leq \mathbb{E}^{0}_{t}\Big[-\alpha_{t,T}\text{exp}\Big(-\gamma\big(3(\mathcal{T}-\Gamma^l-\Gamma^d)\overline{q}^{2}\big(\sum_{i\in \{a,b\}}\sum_{j\in \{l,d\}}N_{T}^{i,j}-N_{t}^{i,j}\big)+\int_{t}^{T}Q_{u}\mathrm{d}\tilde{S}_{u}\big)\Big)\exp(-\gamma \xi)\Big]<0.
\end{align*}

\textbf{Step 3:} Let $Y$ be  the  process  defined for any $t\in [0,T]$ by $V_{t}=-\exp(-\gamma Y_{t})$. As $A^{\mathcal{L},d}$ is a predictable point process and the jumps of $N^{i,j,k}, i\in \{a,b\}, j\in \{l,d\}, k\in\mathcal{V}^j$ are totally inaccessible stopping times under $\mathbb{P}^{0}$, we have $\big\langle N^{i,j,k},A^{\mathcal{L},d}\big\rangle_t=0$ a.s. We obtain
\begin{align*}
Y_{T}=\xi \text{ and } \mathrm{d}Y_{t}=\Big(\sum_{\substack{i\in\{a,b\}\\ j\in\{l,d\}}}\sum_{k\in \mathcal{V}^j}Z_{t}^{i,j,k}\mathrm{d}N_{t}^{i,j,k}\Big)+Z_{t}^{\tilde{S}}\mathrm{d}\tilde{S}_{t}-dI_{t}-d\tilde{A}_{t}^{d}.
\end{align*}
Ito's formula yields to
\begin{align*}
& Z_{t}^{a,l,k}:=-\frac{1}{\gamma}\text{log}\Big(1+\frac{\tilde{Z}_{t}^{\mathcal{L},a,l,k}}{U_{t^{-}}^{\mathcal{L}}}\Big)-\ell_{t}^{a,l}\Big(\frac{\mathcal{T}}{2} +\Gamma^{l}Q_{t^-}\Big)+\Gamma^{l}k^{2}, \\
& Z_{t}^{b,l,k}:=-\frac{1}{\gamma}\text{log}\Big(1+\frac{\tilde{Z}_{t}^{\mathcal{L},b,l,k}}{U_{t^{-}}^{\mathcal{L}}}\Big)-\ell_{t}^{b,l}\Big(\frac{\mathcal{T}}{2} -\Gamma^{l}Q_{t^-}\Big)+\Gamma^{l}k^{2}, \\
& Z_{t}^{a,d,k}:=-\frac{1}{\gamma}\text{log}\Big(1+\frac{\tilde{Z}_{t}^{\mathcal{L},a,d,k}}{U_{t^{-}}^{\mathcal{L}}}\Big)-\ell_{t}^{a,d}\Big(\frac{\mathcal{T}}{2}\mathbf{1}_{\nu_{t}^{a}=0}+\Gamma^{d}Q_{t^-}\Big)+\Gamma^{d}k^{2}, \\
& Z_{t}^{b,d,k}:=-\frac{1}{\gamma}\text{log}\Big(1+\frac{\tilde{Z}_{t}^{\mathcal{L},b,d,k}}{U_{t^{-}}^{\mathcal{L}}}\Big)-\ell_{t}^{b,d}\Big(\frac{\mathcal{T}}{2}\mathbf{1}_{\nu_{t}^{b}=0}-\Gamma^{d}Q_{t^-}\Big)+\Gamma^{d}k^{2}, \\
& Z_{t}^{\tilde{S}}:=-\frac{\tilde{Z}_{t}^{\mathcal{L},S}}{\gamma U_{t^{-}}^{\mathcal{L}}}-Q_{t^{-}}, \\
& I_{t}:=\int_{0}^{t}\Big(\overline{h}(\mathcal{L}_{r},Z_{r},Q_{r})\mathrm{d}r -\frac{1}{\gamma U_{r}^{\mathcal{L}}}dA_{r}^{\mathcal{L},c}\Big), \\
& \overline{h}(\mathcal{L},Z_{t},Q_{t}):=h(\mathcal{L},Z_{t},Q_{t})-\frac{1}{2}\gamma\sigma^{2}(Z_{t}^{\tilde{S}})^{2}, \\
& \tilde{A}_{t}^{d}:=\frac{1}{\gamma}\sum_{s\leq t}\text{log}\Big(1-\frac{\Delta A_{t}^{\mathcal{L},d}}{U_{t^{-}}^{\mathcal{L}}}\Big).
\end{align*}
In particular, the last relation between $\tilde{A}^{d}$ and $A^{\mathcal{L},d}$ shows that $\Delta a_{t}\geq 0$ is independent of $\mathcal{L}\in \mathcal{A}$, with  $a_t=-\frac{A_{t}^{\mathcal{L},d}}{U_{t^{-}}^{\mathcal{L}}}$ and abusing notations slightly, $\Delta a_t = -\frac{\Delta A_{t}^{\mathcal{L},d}}{U_{t^{-}}^{\mathcal{L}}}$.  \\

In order to complete the proof, we argue in the subsequent steps that $Z \in \mathcal{Z}$ and that, for $t\in [0,T]$, $A_{t}^{\mathcal{L},d}=-\sum_{s\leq t}U_{s^{-}}^{\mathcal{L}}\Delta a_{s}=0$ so that $\tilde{A}_{t}^{d}=0$ and $I_{t}=\int_{0}^{t}\overline{H}(Z_{r},Q_{r})\mathrm{d}r$ where 

\begin{align*}
\overline{H}(Z_{t},Q_{t})=H(Z_{t},Q_{t})-\frac{1}{2}\gamma\sigma^{2}(Z_{t}^{\tilde{S}})^{2}.
\end{align*}

\textbf{Step 4:} Since $V_{T}=-1$, we get that
\begin{align*}
0&=\sup_{\mathcal{L}\in \mathcal{A}}\mathbb{E}^{\mathcal{L}}[U_{T}^{\mathcal{L}}]-V_{0}\\
&=\sup_{\mathcal{L}\in \mathcal{A}}\mathbb{E}^{\mathcal{L}}[U_{T}^{\mathcal{L}}-M_{T}^{\mathcal{L}}]\\
&=\gamma\sup_{\mathcal{L}\in \mathcal{A}}\mathbb{E}^{0}\Big[L_{T}^{\mathcal{L}}\int_{0}^{T}U_{r^{-}}^{\mathcal{L}}(\mathrm{d}I_{r}-\overline{h}(\mathcal{L},Z_{r},Q_{r})\mathrm{d}r+\frac{\mathrm{d}a_{r}}{\gamma})\Big]. 
\end{align*}

Moreover, the controls being uniformly bounded, we have
\begin{align*}
U_{t}^{\mathcal{L}}\leq -\beta_{t}:=V_{t}\text{exp}\Big(-\gamma\big(3(\mathcal{T}-\Gamma^l-\Gamma^d)\overline{q}^{2}(\sum_{i\in \{a,b\}}\sum_{j\in \{l,d\}}N_{t}^{i,j})+\int_{0}^{t}Q_{r}\mathrm{d}\tilde{S}_{r}\big)\Big)<0.
\end{align*}
Then, using $A^{\mathcal{L},d}\geq 0, U^{\mathcal{L}}\leq 0$ and $\mathrm{d}I_{t}-\overline{h}(\mathcal{L},Z_{t},Q_{t})\mathrm{d}t\geq 0$, obtain \begin{align*}
0 &\leq \sup_{\mathcal{L}\in \mathcal{A}}\mathbb{E}^{0}\Big[\alpha_{0,T}\int_{0}^{T}-\beta_{r^{-}}\big(\mathrm{d}I_{r}-\overline{h}(\mathcal{L},Z_{r},Q_{r})\mathrm{d}r + \frac{\mathrm{d}a_{r}}{\gamma}\big)\Big]\\
& =-\mathbb{E}^{0}\Big[\alpha_{0,T}\int_{0}^{T}\beta_{r^{-}}\big(\mathrm{d}I_{r}-\overline{H}(Z_{r},Q_{r})\mathrm{d}r + \frac{\mathrm{d}a_{r}}{\gamma}\big)\Big].
\end{align*}

The quantities $\alpha_{0,T}\int_{0}^{T}\beta_{r^{-}}(dI_{r}-\overline{H}(Z_{r},Q_{r}))\mathrm{d}r$ and $\alpha_{0,T}\int_{0}^{T}\beta_{r^{-}}\frac{da_{r}}{\gamma}$ being non-negative random variables, the result follows. \\

Moreover, if $\mathcal{L}$ is such that for any $(z,q)\in\mathbb{R}^{2(\#\mathcal{V}^l+\#\mathcal{V}^d)} \times \mathbb{N}$ we have $h(\mathcal{L},z,q)=H(z,q)$, then
\begin{align*}
    \int_{0}^{T}U_{r^{-}}^{\mathcal{L}}\big(dI_{r}-\overline{h}(\mathcal{L},Z_{r},Q_{r})\big)\mathrm{d}r=0.
\end{align*}
Therefore, $\sup_{\mathcal{L}\in\mathcal{A}}\mathbb{E}^{\mathcal{L}}[U_{T}^{\mathcal{L}}]=V_0$ which implies that \eqref{OC} is satisfied. Conversely, if \eqref{OC} is satisfied, the supremum is directly attained. This provides the inclusion $\mathcal{C}\supset \Xi$. \\

\textbf{Step 5:} As $\xi$ satisfies Conditions \eqref{Integrability market maker} and \eqref{Integrability Principal}, to prove that $Z\in\mathcal{Z}$ it is enough to show that for some $p > 0$
\begin{align*}
\sup_{\mathcal{L}\in \mathcal{A}}\sup_{t\in [0,T]}\mathbb{E}^{\mathcal{L}}\Big[\exp\big(-\gamma(p+1)Y_{t}\big)\Big]<+\infty.
\end{align*}
Using Hölder inequality together with the boundedness of the intensities of the $N^{i,j,k}$, we have that $\sup_{\mathcal{L}\in \mathcal{A}}\mathbb{E}^{\mathcal{L}}[|U_{T}^{\mathcal{L}}|^{p'+1}]<+\infty$ for some $p'>0$. Thus
\begin{align*}
\sup_{\mathcal{L}\in \mathcal{A}}\sup_{t\in [0,T]}\mathbb{E}^{\mathcal{L}}[|U_{t}^{\mathcal{L}}|^{p'+1}]=\sup_{\mathcal{L}\in \mathcal{A}}\mathbb{E}^{\mathcal{L}}[|U_{T}^{\mathcal{L}}|^{p'+1}]<+\infty,
\end{align*}

because $U^{\mathcal{L}}$ is a $\mathbb{P}^{\mathcal{L}}$-negative supermartingale. The conclusion follows using again Hölder inequality, the uniform boundedness of the intensities of the $N^{i,j}$ and the fact that
\begin{align*}
& \exp(-\gamma Y_{t})=U^{\mathcal{L}}_{t}\text{exp}\bigg(\gamma\Big(\int_{0}^{t} \sum_{i\in \{a,b\}}\Big(\frac{\mathcal{T }}{2}\ell_{u}^{i,l}\mathrm{d}N_{u}^{i,l} +\sum_{\kappa\in K}\phi^{lat}(\kappa)\ell_{u}^{i,d}\mathrm{d}N_{u}^{i,d,\kappa}\Big)+Q_{u}\mathrm{d}S_{u}+\mathrm{d}\big[ Q_{\cdot},S_{\cdot}\big]_u\Big)\bigg).
\end{align*}
Consequently, $\mathcal C\subset \Xi$ and using Step 4 we finally get $\mathcal C=\Xi$. \\

\textbf{Step 6:} We prove here uniqueness of the representation. Let $(Y_{0},Z),(Y_{0}^{'},Z^{'})\in \mathbb{R}\times\mathcal{Z}$ be such that $\xi=Y_{T}^{Y_{0},Z}=Y_{T}^{Y_{0}^{'},Z^{'}}$. By following the lines of the verification argument in second part of the proof of the theorem, we obtain the equality $Y_{t}^{Y_{0},Z}=Y_{t}^{Y_{0}^{'},Z^{'}}$ using the fact that the
value of the continuation utility of the market maker satisfies
\begin{align*}
-\exp(-\gamma Y_{t}^{Y_{0},Z})=-\exp(-\gamma Y_{t}^{Y_{0}^{'},Z^{'}})= \underset{\mathcal{L}\in \mathcal{A}}{\text{ess sup }}\mathbb{E}_{t}^{\mathcal{L}}\Big[-\exp\big(-\gamma (PL_{T}^{\mathcal{L}}-PL_{t}^{\mathcal{L}}+\xi)\big)\Big].
\end{align*}

This in turn implies that $Z_{t}^{i,j,k}\mathrm{d}N_{t}^{i,j,k}=Z_{t}^{'i,j,k}\mathrm{d}N_{t}^{i,j,k}$ and $Z_{t}^{\tilde{S}}\sigma^{2}\mathrm{d}t=Z_{t}^{',S}\sigma^{2}\mathrm{d}t=\mathrm{d}\langle Y,S\rangle_{t}, t\in [0,T]$. Thus $(Y_{0},Z)=(Y_{0}^{'},Z^{'})$. \\

We now prove the second part of Theorem \ref{Theorem Market Maker}. Let $\xi=Y_{T}^{Y_{0},Z}$ with $(Y_{0},Z)\in \mathbb{R}\times\mathcal{Z}$. We  first show
that for an arbitrary set of controls $\mathcal{L} \in \mathcal{A}$ we have $J_0^{\text{MM}}(\mathcal{L},\xi)\leq -\exp(-\gamma Y_{0})$ where we recall that  $J_0^{\text{MM}}(\mathcal{L},\xi)$ is such that $V^{\text{MM}}_0(\xi)=\sup_{\mathcal{L}\in \mathcal{A}}J_0^{\text{MM}}(\mathcal{L},\xi)$. Then we will see that this inequality is in fact an equality when the corresponding Hamiltonian $h(\mathcal{L},z,q)$ is maximized. Let us write
\begin{align*}
\overline{Y}_{t}:=Y_{t}^{Y_{0},Z}&+\int_{0}^{t}\frac{\mathcal{T }}{2}\ell_{u}^{a,l}\mathrm{d}N_{u}^{a,l}+\int_{0}^{t}\frac{\mathcal{T }}{2}\ell_{u}^{b,l}\mathrm{d}N_{u}^{b,l}+\int_{0}^{t}Q_{u}\mathrm{d}S_{u}+\mathrm{d}\big[ Q_{\cdot},S_{\cdot}\big]_u \\
&+ \int_{0}^{t}\frac{\mathcal{T }}{2}\ell_{u}^{a,d}\mathrm{d}N_{u}^{a,d,\text{lat}}+\int_{0}^{t}\frac{\mathcal{T}}{2}\ell_{u}^{b,d}\mathrm{d}N_{u}^{b,d,\text{lat}},
\end{align*}
with $t\in [0,T]$. An application of Ito’s formula leads to 
\begin{align*}
\mathrm{d}\big(\exp(-\gamma\overline{Y}_{t})\big)&=\gamma\exp(-\gamma\overline{Y}_{t^{-}})\Bigg(-(Q_{t}+Z_{t}^{\tilde{S}})\mathrm{d}\tilde{S}_{t}+(H(Z_{t},Q_{t})-h(\mathcal{L},Z_{t},Q_{t}))\mathrm{d}t \\
&- \sum_{(k^l,k^d)\in\mathcal{V}^l\times\mathcal{V}^d}\sum_{i\in \{a,b\}}\gamma^{-1}\Bigg(\bigg(1-\text{exp}\Big(-\gamma\big(Z_{t}^{i,l,k^l}+\ell_{t}^{i,l}(\frac{\mathcal{T }}{2}+\Gamma^{l}(\phi(i) Q_{t^-}-\ell_{t}^{i,l})) \big)\Big)\bigg)\mathrm{d}N_{t}^{\mathcal{L},i,l,k^l}\\
&\!-\!\sum_{\kappa\in K}\!\bigg(1\!-\!\exp\Big(\!\!-\!\gamma\big(Z_{t}^{i,d,k^d}\!\!+\!\ell_{t}^{i,d}\!\big(\frac{\mathcal{T}}{2}\! \phi^{\text{lat}}(\kappa)\!+\!\Gamma^{d}(\phi(i)Q_{t^-}\!-\!\ell_{t}^{i,d})\big)\!\big)\!\Big)\!\bigg)\!\phi_{t}^{d}(i,\kappa)\mathrm{d}N_{t}^{\mathcal{L},i,d,k^d}\Bigg)\Bigg).
\end{align*}
Therefore $\exp(-\gamma\overline{Y}_{.})$ is a $\mathbb{P}^{\mathcal{L}}$-local submartingale.  Thanks to Condition \eqref{Strong integrability MM}, the uniform boundedness of the intensities of the $N^{i,j,k}$, $i\in \{a,b\}, j\in \{l,d\},k\in\mathcal{V}^j$ and Hölder inequality, $\exp(-\gamma\overline{Y}_{\cdot})$ is uniformly integrable and hence a true submartingale. Doob-Meyer decomposition gives us that
\begin{align*}
\int_{0}^{\cdot}\gamma\exp(-\gamma\overline{Y}_{t^{-}})\Bigg(&\!-\!(Q_{t}+Z_{t}^{\tilde{S}})\mathrm{d}\tilde{S}_{t} \\
&-\!\!\!\!\!\!\!\!\!\sum_{(k^l,k^d)\in\mathcal{V}^l\times\mathcal{V}^d}\sum_{i\in \{a,b\}}\gamma^{-1}\Bigg(\bigg(1-\text{exp}\Big(-\gamma\big(Z_{t}^{i,l,k^l}+\ell_{t}^{i,l}(\frac{\mathcal{T }}{2}+\Gamma^{l}(\phi(i) Q_{t^-}-\ell_{t}^{i,l})) \big)\Big)\bigg)\mathrm{d}N_{t}^{\mathcal{L},i,l,k^l}\\
&\!-\!\sum_{\kappa\in K}\!\bigg(1\!-\!\exp\Big(\!-\!\gamma\big(\!Z_{t}^{i,d,k^d}\!+\!\ell_{t}^{i,d}\big(\frac{\mathcal{T}}{2} \phi^{\text{lat}}(\kappa)\!+\!\Gamma^{d}(\phi(i)Q_{t^-}-\ell_{t}^{i,d})\big)\!\big)\!\Big)\!\bigg)\phi_{t}^{d}(i,\kappa)\mathrm{d}N_{t}^{\mathcal{L},i,d,k^d}\Bigg)\Bigg)
\end{align*}
is a true martingale. Thus
\begin{align*}
J_0^{\text{MM}}(\mathcal{L},\xi)&=\mathbb{E}^{\mathcal{L}}\Big[-\exp(-\gamma\overline{Y}_{T})\Big]\\
&=-\exp(-\gamma Y_{0})-\mathbb{E}^{\mathcal{L}}\bigg[\int_{0}^{T}\gamma\exp(-\gamma\overline{Y}_{t^{-}})\big(H(Z_{t},Q_{t})-h(\mathcal{L},Z_{t},Q_{t})\big)\mathrm{d}t\bigg] \\
&\leq -\exp(-\gamma Y_{0}).
\end{align*}
In addition to this, the previous inequality becomes an equality if and only if $\mathcal{L}$ is chosen as the maximizer of the Hamiltonian $h$. In that case, $J_{\text{MM}}(\mathcal{L},\xi)=-\exp(-\gamma Y_{0})$. Finally we have that $V^{\text{MM}}_0(\xi)=-\exp(-\gamma Y_{0})$ with optimal response $(\mathcal{L}^{\star}_{t})_{t\in [0,T]}$ defined by \eqref{OC}.

\subsection{Proof of Theorem \ref{Main Theorem 2}}\label{proof:visco}
We recall that, by \cite[Corollary 1.4.2]{bouchard2007introduction}, the PDE \eqref{HJB Equation Principal First Form} admits a unique continuous viscosity solution denoted by $v$. \\

Let $(t_0,\tilde{s}_0,\bar{n}_0,n_0,y_0)\in \mathfrak D$ where $\mathfrak D=[0,T]\times \mathbb R\times \mathbb N^{2(\#\mathcal{V}^l+\#\mathcal{V}^d)}\times \mathbb N^{2(\#\mathcal{V}^l+\#\mathcal{V}^d)} \times \mathbb R$. We consider a test function $\Phi:\mathfrak D\rightarrow\mathbb{R}$ continuously differentiable in time, twice continuously differentiable with respect to $s$ and $y$ and continuous with respect to $\bar n$ and $n$, such that
\begin{align*}
    0 &=u(t_0,\tilde{s}_0,\bar{n}_0,n_0,y_0)-\Phi(t_0,\tilde{s}_0,\bar{n}_0,n_0,y_0)\\
    & =\max_{(t,\tilde{s},\bar{n},n,y)\in\mathfrak D}\exp\!\big(\!\!-\eta(\!\sum_{\substack{i\in\{a,b\}\\j\in\{l,d\}}}\sum_{k\in\mathcal{V}^j}\!\! c^j \bar{n}^{i,j,k}\!-\! y)\big)\Big(v(t,q)-\Phi(t,\tilde{s},\bar{n},n,y)\exp\big(\eta(\sum_{\substack{i\in\{a,b\}\\j\in\{l,d\}}}\sum_{k\in\mathcal{V}^j}c^j \bar{n}^{i,j,k}-y)\big)\Big).
\end{align*}
Therefore for all $(t,\tilde{s},\bar{n},n,y)\in\mathfrak D$
\begin{align*}
    0 \geq v(t,q)-\Phi(t,\tilde{s},\bar{n},n,y)\exp\big(\eta(\sum_{\substack{i\in\{a,b\}\\j\in\{l,d\}}}\sum_{k\in\mathcal{V}^j}c^j \bar{n}^{i,j,k}-y)\big),
\end{align*}
with equality at $(t_0,\tilde{s}_0,\bar{n}_0,n_0,y_0)$. Thus
\begin{align*}
    0&=v(t_0,q_0)-\Psi(t_0,\tilde{s}_0,\bar{n}_0,n_0,y_0)\\
    & =\max_{(t,\bar{n})\in\mathfrak D}\Big(v(t,q)-\Psi(t,\bar{n})\Big),
\end{align*}
where 
\begin{align*}
\Psi(t,\bar{n}):=\Phi(t,\tilde{s}_0,\bar{n},n_0,y_0)\exp\big(\eta(\sum_{\substack{i\in\{a,b\}\\j\in\{l,d\}}}\sum_{k\in\mathcal{V}^j}c^j \bar{n}^{i,j,k}-y_0)\big).
\end{align*}
As $v$ is the unique viscosity solution of \eqref{HJB Equation Principal First Form}, it is in particular a subsolution. Thus, for any $z\in \mathbb R^{2(\#\mathcal{V}^l+\#\mathcal{V}^d)+1}$, $\Psi$ satisfies
\[   0\geq \partial_{t}\Psi(t_0,\bar n_0) +\mathcal{U}\big(z,q_0,\mathcal{L}^\star(z,q_0),\Psi(t_0,\cdot)\big),\]
with $q_0:= \sum_{j\in\{l,d\}}\sum_{k\in\mathcal{V}^j}(\bar n_0^{b,j,k} - \bar n_0^{a,j,k})$, $\mathcal{U}$ is defined by \eqref{hamiltonian} and $\mathcal{L}^\star$ is defined in Theorem \ref{Theorem Market Maker}. After computations, we deduce that
\begin{align*}
 0\geq& \partial_t \Psi(t_0,\bar{n}_0) +\Psi(t_0,\bar{n}_0)\Big(\frac{\eta}{2}\sigma^{2}\gamma\big(z^{\tilde{S}}+q_0\big)^{2}+\frac{\eta^{2}\sigma^{2}}{2}\big(z^{\tilde{S}}\big)^{2}\Big) + \sum_{\substack{i\in\{a,b\}\\j\in\{l,d\}}}\sum_{k\in\mathcal{V}^j} \lambda^{\mathcal{L}^\star,i,j,k} \\ \times&\bigg(\exp\big(\eta(z^{i,j,k}-k c^{j})\big)\Phi(t_0,\tilde{s}_0,\bar{n}^{i,j,k}_0+k,\bar{n}^{-(i,j,k)}_0,n_0,y_0)\exp\big(\eta(\sum_{\substack{i\in\{a,b\}\\j\in\{l,d\}}}\sum_{k\in\mathcal{V}^j}c^j \bar{n}_0^{i,j,k}-y_0)\big)\\
 & -\Psi(t_0,\tilde{s}_0,\bar{n}_0,n_0,y_0)\big(1+\eta \mathcal{E}(z^{i,j,k},\ell^{\star i,j}(z,q_0))\big)\bigg).
\end{align*}
Dividing on both sides of the equation by $\exp\big(\eta(\sum_{\substack{i\in\{a,b\}\\j\in\{l,d\}}}\sum_{k\in\mathcal{V}^j}c^j \bar{n}_0^{i,j,k}-y_0)\big)> 0$, we obtain
\begin{align*}
 & 0\geq \partial_t \Phi (t_0,\tilde{s}_0,\bar{n}_0,n_0,y_0)+\Phi_0\Big(\frac{\eta}{2}\sigma^{2}\gamma\big(z^{\tilde{S}}+q_0\big)^{2}+\frac{\eta^{2}\sigma^{2}}{2}\big(z^{\tilde{S}}\big)^{2}\Big)+ \sum_{\substack{i\in\{a,b\}\\j\in\{l,d\}}}\sum_{k\in\mathcal{V}^j} \lambda^{\mathcal{L}^\star,i,j,k} \\
  &\hspace{1em}\times\bigg(\exp\big(\eta(z^{i,j,k}-kc^{j})\big)\Phi(t_0,\tilde{s}_0,\bar{n}^{i,j,k}_0+k,\bar{n}^{-(i,j)}_0\!\!,n_0,y_0)\!-\!\Phi_0\!\big(1+\eta \mathcal{E}(z^{i,j,k}\!,\!\ell^{\star i,j}(z,q_0))\big)\!\bigg),
\end{align*}
 where $\Phi_0:=\Phi(t_0,\tilde{s}_0,\bar{n}_0,n_0,y_0)$. Therefore, $u$ is a viscosity subsolution of \eqref{HJB Principal before change variable}. A similar argument holds to prove that $u$ is also a viscosity supersolution of \eqref{HJB Principal before change variable}. Consequently, $u$ is a viscosity solution of \eqref{HJB Principal before change variable}. The uniqueness of $u$ follows from an application of \cite[Theorem II.3]{lions1983hamilton}, together with the continuity of $v$. Thus, we deduce that $v_0^E = u(0, \tilde{S}_0, \bar{N}_0, N_0, Y_0) = v(0, Q_0)$.
\endgroup

\bibliographystyle{abbrv}
\bibliography{bibliographyBMR.bib}

\end{document}